\def \Mpc {h^{-1}{\rm Mpc}}
\def \kpc {h^{-1}{\rm kpc}}

\def \farcm{\hbox{$.\!\!^{\prime}$}}

\def \kms {{\rm km~s}^{-1}}
\def \msun {h^{-1} M_\odot}
\def \beqn {\begin{equation}}
\def \eeqn {\end{equation}}
\documentstyle[12pt,aaspp4]{article}

\begin{document}

\title{The Infall Region of Abell 576: \\Independent Mass and Light
Profiles}
%\shorttitle{A576 Infall Region}
%\shortauthors{Rines et al.}

\author{Kenneth Rines and Margaret J. Geller}
\affil{Harvard-Smithsonian Center for Astrophysics, 60 Garden St,
Cambridge, MA 02138 \\ krines, mgeller@cfa.harvard.edu}

\author{Antonaldo Diaferio}
\affil{Universit\`a degli Studi di Torino,
Dipartimento di Fisica Generale ``Amedeo Avogadro'',
Torino,
Italy}

\author{Joseph J. Mohr\altaffilmark{1}}
\affil{Department of Astronomy and
Astrophysics, University of Chicago, Chicago, IL
60637}

\author{Gary A. Wegner}
\affil{Department of Physics and Astronomy, Dartmouth College,
Hanover, NH 03755}
\authoremail{krines@cfa.harvard.edu}

\altaffiltext{1}{Chandra Fellow}

\begin{abstract}
We describe observations of the nearby ($cz=11,487~\kms$) cluster of
galaxies Abell 576 beyond the 
virial radius and into the infall region where galaxies are on their
first or second pass through the cluster. Using 1057 redshifts, we
use the infall pattern in redshift space to determine the mass profile
of A576 to a radius of  
$\sim 4~\Mpc$. This mass estimation technique makes no assumptions
about the equilibrium state 
of the cluster. Within $\sim 1~\Mpc$, the mass profile we derive exceeds that
determined from X-ray observations by a factor of $\sim 2.5$. At
$\sim 2.5~\Mpc$, however, the mass profile agrees with virial mass 
estimates. Our mass profile is consistent with a Navarro,
Frenk, \& White (1997) or Hernquist (1990) profile, but it is inconsistent
with an isothermal sphere.   

R-band images of a $3^\circ \times 3^\circ$ region centered on the
cluster allow an independent determination of the cluster light
profile. We calculate the integrated
mass-to-light ratio as a function of cluster radius; it
decreases smoothly from the core to  $M/L_R \sim 300 h$ at
$\sim 4~\Mpc$. The differential $M/L_R$ profile decreases more
steeply; we find $\delta M/\delta L_R \sim 100 h$ at $\sim 4~\Mpc$, in
good agreement with the mass-to-light ratios of individual
galaxies. If the behavior of $M/L_R$ in A576 is general, $\Omega _m
\lesssim 0.4$ at 95\% confidence.  

For a Hernquist model, the best-fit mass profiles differ from the
observed surface number density of galaxies; 
the galaxies have a  larger scale radius than the
mass. This result is consistent with the
centrally peaked $M/L_R$ profile. Similarly, the scale radius of the
light profile is larger than that of the mass profile. We discuss some
potential systematic effects; none can easily reconcile
our results with a constant mass-to-light ratio.

\end{abstract}


\section{Introduction}

Clusters of galaxies are important probes of the
distribution of both matter and light on intermediate scales
($0.1-10~\Mpc$). They 
are also interesting laboratories for studying the 
effects of local environment on cluster galaxies. Most studies of
clusters concentrate on the central regions where the cluster is probably
in equilibrium. Studies of the galaxy
distribution on larger scales tend to focus either on general
properties of 
large-scale structure (e.g., \cite{dgh86};
\cite{ga}) or on individual superclusters (e.g., Giovanelli \& Haynes
1985). There are relatively few examinations (\cite{rg89};
\cite{lilje}; \cite{vanh}; \cite{vanh2}; \cite{praton}; \cite{dg};
\cite{vedel}; \cite{ellingson}) of the infall regions of 
clusters. In these regions, the galaxies are falling into the 
gravitational potential well of the cluster, but they have not yet reached
equilibrium. Many, perhaps most, of the galaxies in this region are
on their first orbit of the cluster. They populate a regime
between that of relaxed cluster cores and the surrounding large-scale
structure where the transition from linear to non-linear clustering
occurs. 

Here we use 1057 redshifts and photometry of 2118 galaxies in the
infall region of Abell 576 ($cz=11,487~\kms$) to address an unresolved 
problem in astrophysics: the relative distribution of mass and light
on large scales. Zwicky (1933; 1937) originally found that the mass of 
the Coma cluster greatly exceeds the sum of the masses
of the stars. More recently, calculations of mass-to-light ratios
for galaxy clusters yield values of several hundred in solar units
(\cite{d78}; \cite{fg79};
\cite{enacs98}). Girardi et al.~(2000) calculated
mass-to-light ratios for a large sample of nearby clusters within the
virial radius and obtained typical values of $M/L_{B_j} \sim 220-250
h$ (all $M/L$ values are in solar units, i.e.,
$M_\odot/L_\odot$). David, Jones, \& Forman (1995) find $M/L_V \sim
200-300 h$ for seven groups and clusters using masses calculated
from the observed X-ray emission, and the CNOC survey finds $M/L_r
\sim 290 \pm 60 h$ for a sample of distant clusters
(\cite{c96}). 

The mass density parameter  $\Omega_m$ of the universe can be
estimated by assuming that the universal mass-to-light ratio is equal
to the ratio in rich clusters of galaxies (e.g., Carlberg et
al.~1996). There have been 
few attempts to determine mass-to-light ratios directly at larger
radii (\cite{mw97}; \cite{small98}; \cite{k99}), where clusters are in
neither hydrostatic nor virial equilibrium. Neither X-ray
observations nor virial analysis provide accurate mass
determinations at these large radii. Two methods with particular
promise are weak gravitational lensing (\cite{k99}) and
kinematics of the infall region (Diaferio \& Geller 1997, hereafter
DG; Diaferio 1999, hereafter D99). Kaiser et al.~analyzed the
weak lensing signal from a
supercluster at $z \approx 0.4$ and found no significant evidence for
variations in mass-to-light ratios on scales less than
$\sim 6~\Mpc$. Metzler et al.~(1999), however, show that lensing by
intervening filamentary structures 
probably associated with clusters can result in significant
overestimates of cluster masses. Geller, Diaferio, \& Kurtz (1999,
hereafter GDK) applied the kinematic method of DG to the infall region
of the Coma 
cluster. They successfully reproduced the X-ray derived mass profile and
extended direct determinations of the mass profile to $\sim 10~\Mpc$. 
Adequate photometric data necessary to compute the mass-to-light profile for
this large region around Coma are not yet available.

Here we apply the method of DG to A576. In redshift space, the infall
regions of clusters form a 
characteristic trumpet-shaped pattern. These caustics arise because
galaxies fall into the cluster as the cluster potential
overwhelms the Hubble flow  
(\cite{kais87}; \cite{rg89}). Under simple spherical infall, the
galaxy phase space density becomes infinite at the caustics. DG
analyzed the dynamics of infall regions with numerical simulations and
found that in the outskirts of clusters, random motions due to
substructure and non-radial motions make a substantial contribution to
the amplitude of the  
caustics which delineate the infall regions. DG showed that the
amplitude of the caustics is a measure of the escape velocity from the
cluster; identification of the caustics therefore allows a
determination of the mass profile of the cluster on scales $\lesssim
10\Mpc$.

DG show that nonparametric measurements
of caustics would yield cluster mass profiles accurate to $\sim$30\%
on scales 
$1-10~\Mpc$, if (1) the redshift space coordinates of the dark matter
particles were measurable, and (2) the cluster mass within the virial
radius were known exactly. More realistically, by using
simulated catalogs of galaxies
formed and evolved using semi-analytic procedures within the
dark matter halos of dissipationless N-body simulations
(\cite{kauf98a}), D99 shows that the identification of caustics in the
realistic redshift diagrams of  clusters recovers their mass
profiles within a 
factor of 2 to several Megaparsecs from the cluster center without a
separate determination of the central mass.
When combined with wide-field photometry, this
approach allows a determination of the 
mass-to-light ratio on large scales which is independent of
the assumption that light traces mass.
This method assumes only that galaxies trace the velocity
field. Indeed, simulations suggest that velocity bias, if any, is very
weak on both linear and non-linear scales (\cite{kauf98a};
\cite{dkw99}). Vedel \& Hartwick (1998) used   
simulations to explore an alternative parametric maximum likelihood
analysis of the infall region. Their technique
requires assumptions about the functional forms of the density profile
and the velocity dispersion profile.  

Here, we analyze the caustics within a $\sim 4~\Mpc$ radius to
determine the mass profile of  
Abell 576, an Abell richness class 1 cluster (\cite{abell}) at a
redshift of $z=0.0383$. Mohr et al.~(1996,
hereafter M96) extensively studied the inner square degree of this
cluster. Using
photometric observations of a $3^\circ \times 3^\circ$ region centered
on the cluster, we determine the mass-to-light profile within this range. The
integrated mass-to-light ratio smoothly decreases with radius out to
$\sim 4~\Mpc$. Remarkably, the differential mass-to-light ratio
decreases steeply to values typical of individual galaxy halos.

We describe the photometric
and spectroscopic observations in $\S$ 2. In $\S$ 3, we determine the
amplitude of the caustics and calculate the mass profile. We analyze
the X-ray 
observations in $\S$ 4 and compare the X-ray mass profile to the
infall mass profile. We discuss the contribution of background
galaxies in $\S$ 5. We determine
the surface number density profile in $\S$ 6 and compare it to the
mass distribution. 
We test for luminosity segregation in $\S$ 7 and calculate the
light profile in $\S$ 8.  
We compare the mass and light profiles
of the cluster in $\S 9$. We discuss possible systematic effects in
$\S 10$ and conclude in $\S 11$.

\section{Data}
\subsection{Images}

We obtained ten 200-second R-band images of A576 with the
MOSAIC camera on the 0.9-m telescope at Kitt Peak on 1999 February 13-14;
both nights were photometric. The MOSAIC camera consists of 8 CCD
chips, each with a field of view of 15$' \times 30'$. Our
mosaic thus covers approximately a $3^\circ \times 3^\circ$ region
centered on A576.  Adjacent images overlap by
1$\farcm$0. We imaged the central 
square degree on both nights; the 14 February image is offset
1$\farcm$0 N and 1$\farcm$0 E from the 13 February image. We reduced
the images using standard IRAF procedures in the {\it mscred}
package. We use SExtractor (\cite{sex}) to locate sources in the
images and calculate magnitudes. SExtractor divides the image into
segments and assigns pixels to individual objects to deal with crowded
images. Analysis of simulated CCD images show that SExtractor
accurately recovers total magnitudes of faint galaxies
(\cite{sex}). We use the 
MAG\_BEST magnitudes, which are equivalent to MAG\_AUTO, an
aperture magnitude, unless an object has one or more close neighbors
likely to contaminate the flux by more than 10\%, in which case
MAG\_BEST=MAG\_ISOC, an isophotal magnitude with a correction based on 
a Gaussian light profile. We classify all objects with
{CLASS\_STAR$<$~0.6} as galaxy candidates; we then visually inspect these to
eliminate binary stars and artifacts. We calibrate
the photometry with observations of M67 (\cite{m67b}; \cite{m67a}) and
Landolt (1992) fields. We obtain the color correction term from the
data set for 
the entire observing run and the extinction coefficient from fits to
data on all chips for each night. To allow for possible chip-to-chip 
sensitivity variations, we calculate the zero point separately for
each chip (\cite{warren}).  

We use the IRAF package {\it
apphot} to test the accuracy of the SExtractor magnitudes. Using
curves of growth on the images of a few dozen galaxies covering the
range of magnitudes $m_R = 14$ to $m_R$ = 18, we estimate their total
magnitudes within large apertures (50-150$''$).  We find excellent
agreement ($\lesssim 0.1$ mag) between the IRAF total aperture
magnitudes and the SExtractor MAG\_BEST magnitudes. 

\begin{figure}
\figurenum{1}
\label{figcompmagn1n2}
\plotone{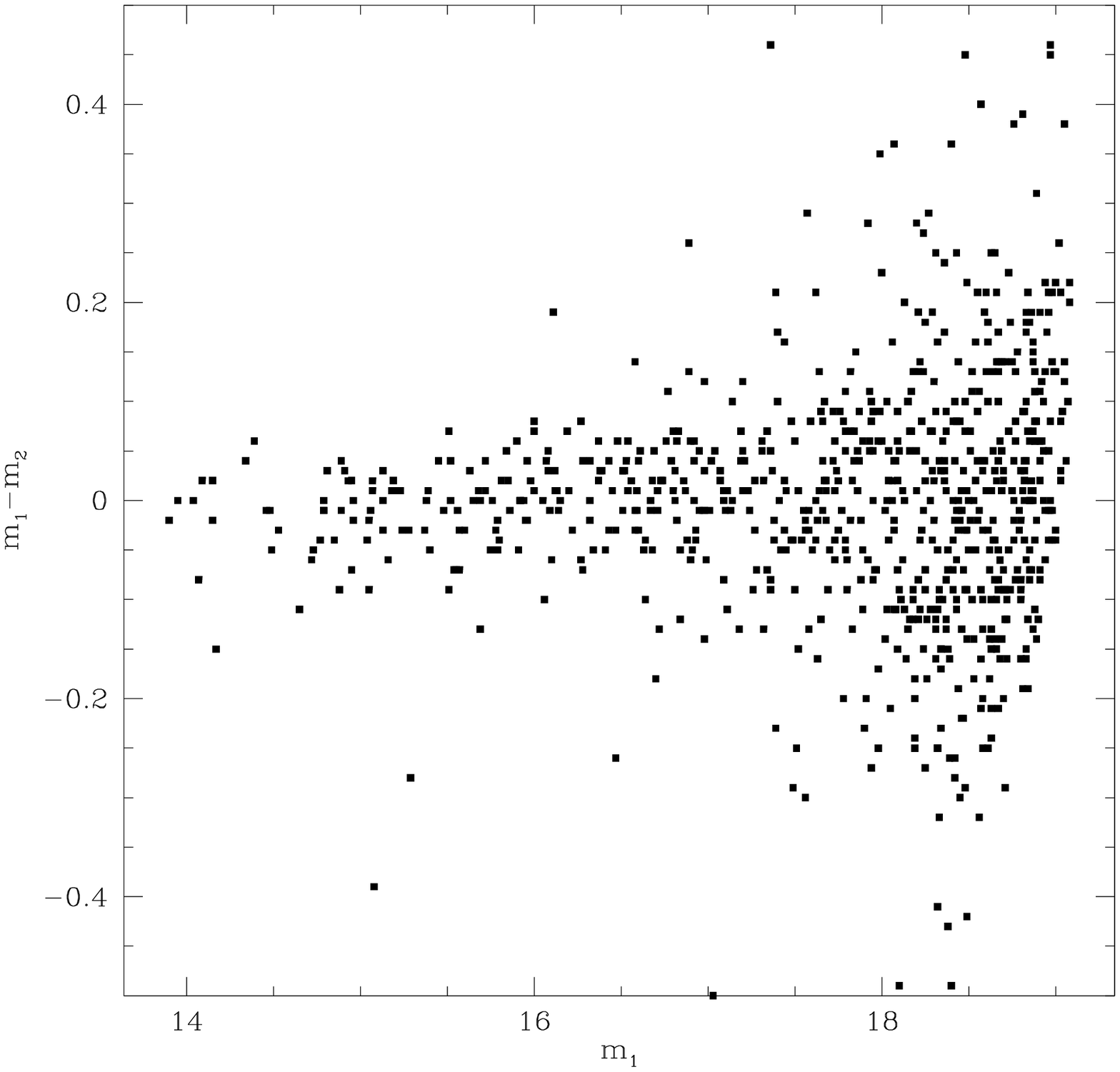} 
\caption{Comparison of magnitudes from the two nights for galaxies in
the central region. For $m_R<18$, the zero-point offset is -0.002 mag
with $\sigma \approx 0.09$ mag. The scatter increases with apparent
magnitude.}
\end{figure}

We restrict our analyses to objects with {MAG\_BEST$<$18.0},
approximately the 
limit of the classification. We use the offset images of the
central region to estimate the consistency of our magnitudes; the
distribution of $m_2-m_1$ ($m_1$ and $m_2$ are the magnitudes
calculated from the first and second night respectively) for galaxies
with $m_R<$18.0 is well represented by a Gaussian 
with a zero point difference of -0.002~mag and $\sigma_{(m_2-m_1)}
\approx 0.09$ mag. The scatter in 
the magnitude determination increases significantly with apparent
magnitude (Figure \ref{figcompmagn1n2}). We thus
estimate that our magnitudes are internally consistent to about
0.06 mag ($\sigma_{m_1}^2 \approx \sigma_{m_2}^2 \approx
{\sigma_{(m_2-m_1)}}^2/2$). In Table 1, we
list the R band magnitudes and their uncertainties (quadrature sums of
the internal SExtractor errors and the 0.06 mag scatter from galaxies
with multiple observations) for galaxies with 
redshifts from FLWO ($\S 2.2$). We also include estimates of Galactic
extinction in the R band ($A_R$) based on the dust maps of Schlegel,
Finkbeiner, \& Davis (1998). Mohr et al.~(2000) will list redshifts
collected by JJM and GW. Magnitudes for these galaxies as well as
those without redshifts are available
upon request from the authors. In the entire survey region, we identify
825 and 2118 galaxies with $m_R<$17.0 and $m_R<$18.0 respectively. 

\begin{figure}
\figurenum{2}
\label{figcompmagm96}
\plottwo{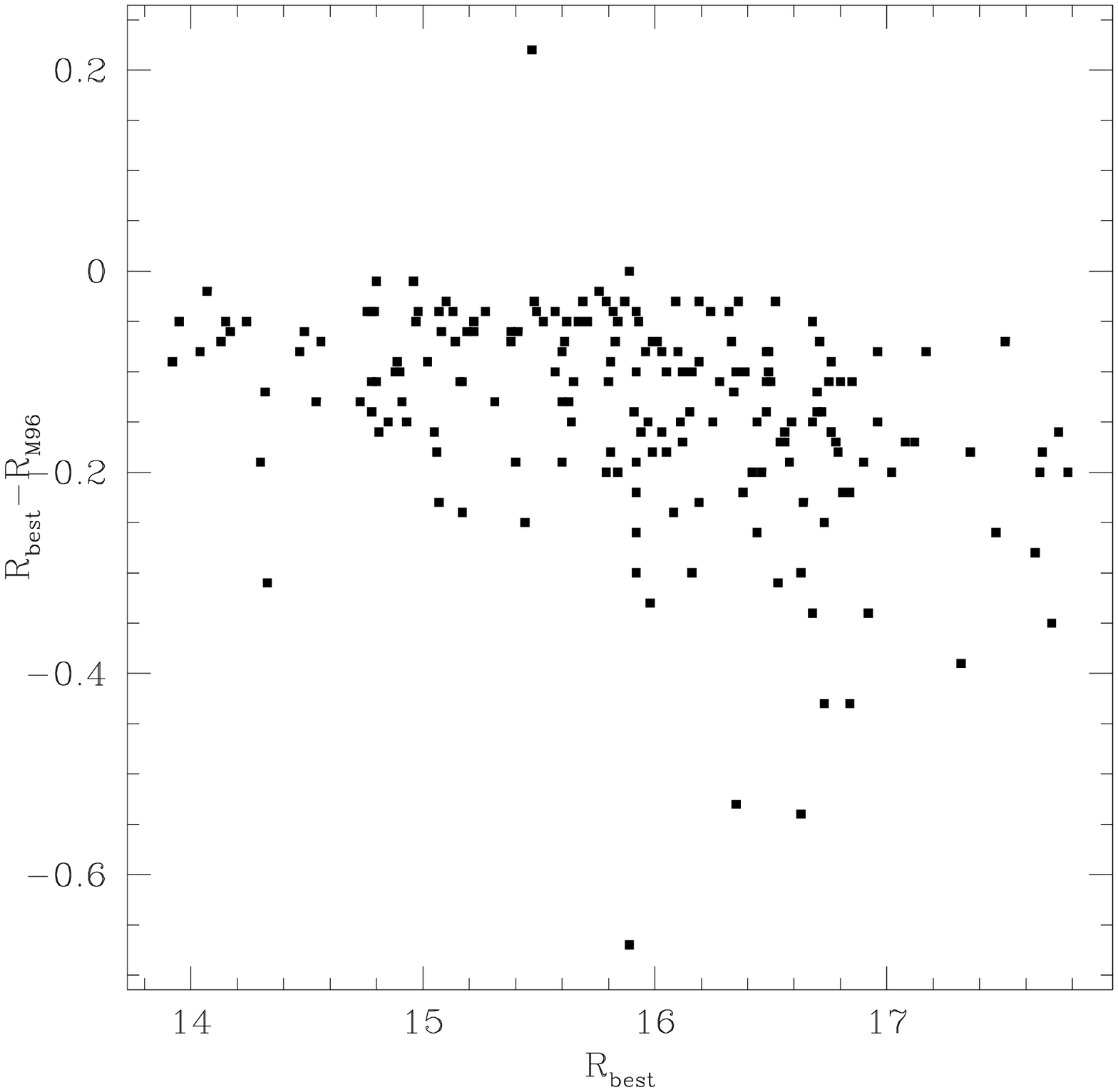}{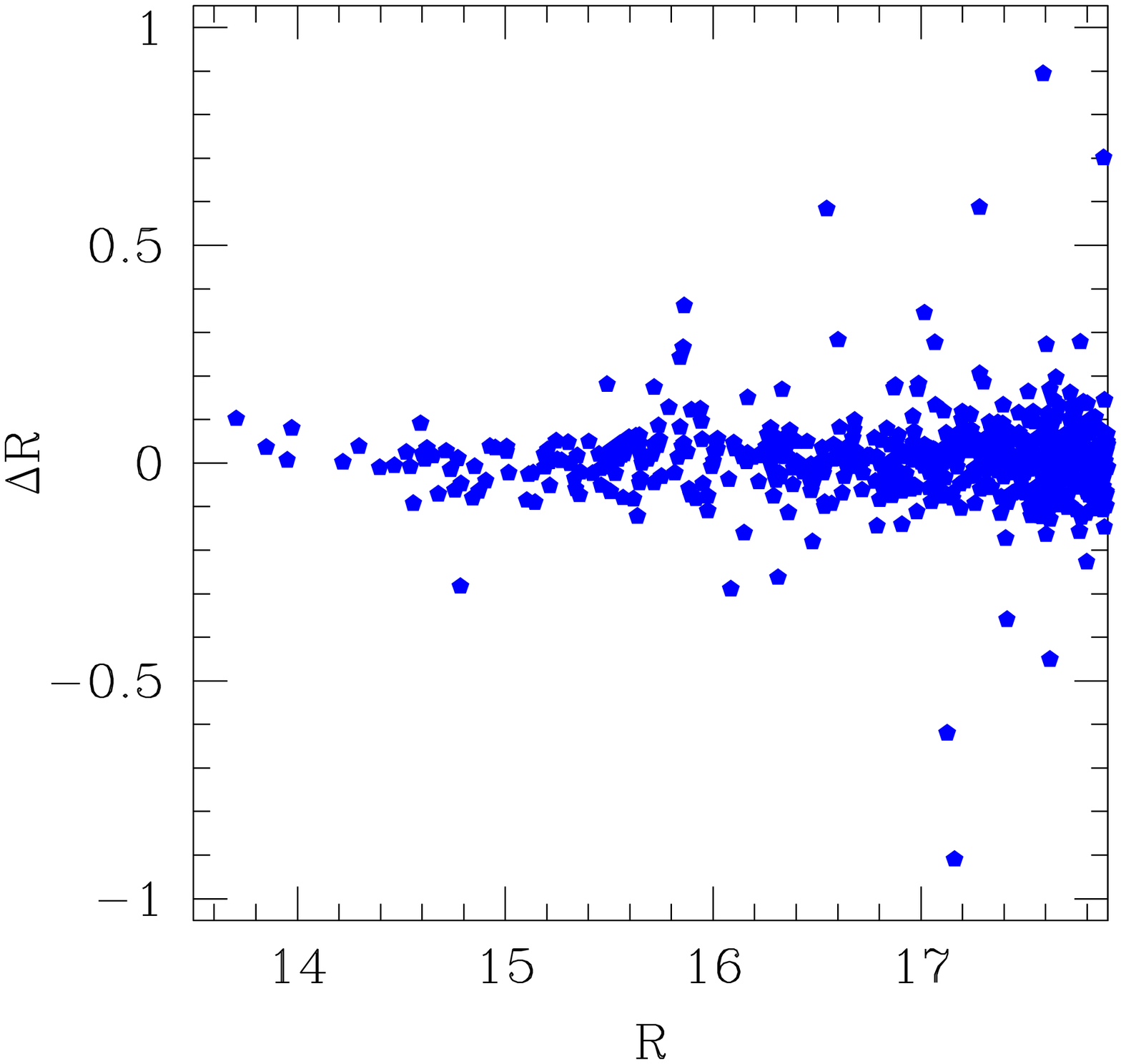} 
\caption{(a) Comparison of our MAG\_BEST magnitudes with those of
M96. Our magnitudes are systematically brighter. (b)
Comparison of our magnitudes with those of M96 after processing both
data sets with SExtractor.} 
\end{figure}

We compare our photometry with that of M96. Figure
\ref{figcompmagm96}a shows the difference in 
magnitude between the two studies as a function of our 
magnitudes. The comparison suggests that our magnitudes are 
systematically brighter by $\sim 0.1$ mag. Further analysis, however,
shows that the differences between our  
magnitudes and M96 can be attributed to differences in the
reduction packages (SExtractor versus FOCAS). For example, FOCAS uses
a global estimate of background counts while SExtractor uses a local
estimate. A complete comparison of the differences between these
packages is beyond the scope of this work. A reanalysis of the M96
data using SExtractor produces excellent agreement (Figure
\ref{figcompmagm96}b) between the two independent photometric data sets 
(\cite{mw00}). 

Out of 823 galaxies with redshifts selected without reference to 
these images ($\S 2.2$), SExtractor misses 28. Visual inspection 
reveals that 17 of these galaxies are in the halos of bright stars;
the rest lie in chip gaps. For the 17 galaxies in stellar halos, we
perform photometry with {\it apphot}. Unfortunately, three of these 
galaxies are located within $\sim 1'.5$ of the X-ray center of the
cluster (a saturated star is $1'.1$ from the X-ray center), which
increases the uncertainty of the central light. We make no correction
for the regions of sky covered by saturated stars; this approach may
lead to a slight underestimate in the light profile.

\subsection{Spectroscopy}

We used the FAST spectrograph 
(\cite{f98}) on the 1.5-m Tillinghast telescope of the Fred Lawrence
Whipple Observatory (FLWO) to obtain 529 spectra of galaxies within $4^\circ$
($\approx 8~\Mpc$) of the center of A576. FAST is a long-slit
spectrograph with a CCD detector. Integration times were
typically 4-20 minutes, with spectral resolution of 6-8 \AA. We
analyzed the spectra at the Telescope Data Center at 
the Smithsonian Astrophysical Observatory using the {\it xcsao} and {\it
emsao} tasks, which are standard
cross-correlation fitting routines (\cite{kurtz}) available in the IRAF
package {\it r2rvsao}. In this technique, template
galaxy spectra are cross-correlated with the observed log-wavelength
binned spectra to
determine the redshift which provides the best fit for a particular
template. Cross-correlation fits with R values of $\gtrsim 4.0$ are
acceptable; most spectra have larger R values
(the goodness of fit increases with R value).

We observed infall galaxy candidates in two campaigns; these occurred
before and after obtaining the MOSAIC images described above. To select
candidates for the first campaign (1999 January-February), we obtained
galaxy positions from digital scans of the POSS I 
plates, using a method developed by Daniel Koranyi (1999) to
discriminate between stars and galaxies. We extracted sources from the
plate scans with SExtractor. We included all objects with
staricity {parameter $<$0.6} and ISO7-to-ISO2 isophotal ratio greater
than 0.25. Comparison with visual inspection
suggests that $\lesssim 1\%$ of galaxies are misclassified as stars
using this criterion. We visually inspected the remaining objects to
eliminate  stars from the sample. For galaxies with $m_R
\lesssim 15.3$, we measured redshifts in magnitude order. Because
the magnitude order was taken from the plate scans, this ordering is
not strictly accurate, and we could not obtain redshifts for some low
surface brightness galaxies. Out of 300 candidates, this campaign
yielded 293 
redshifts (11 candidates were stars, plate flaws, or unobservable;
4 galaxies were serendipitously observed).

In the second campaign (1999 October-2000 February), we selected
candidates from the MOSAIC images. Prior to the second campaign, our
redshift sample was $\sim 95\%$ complete for $m_R<15.3$. In the second
campaign, we observed 236 galaxies in magnitude order as determined
from the MOSAIC images. We list the redshifts and $m_R$ magnitudes
of the galaxies from both campaigns in Table 1. Table
2 lists the positions and 
redshifts of galaxies outside the MOSAIC images.

\begin{figure}
\figurenum{3}
\label{figfracobserved}
\plotone{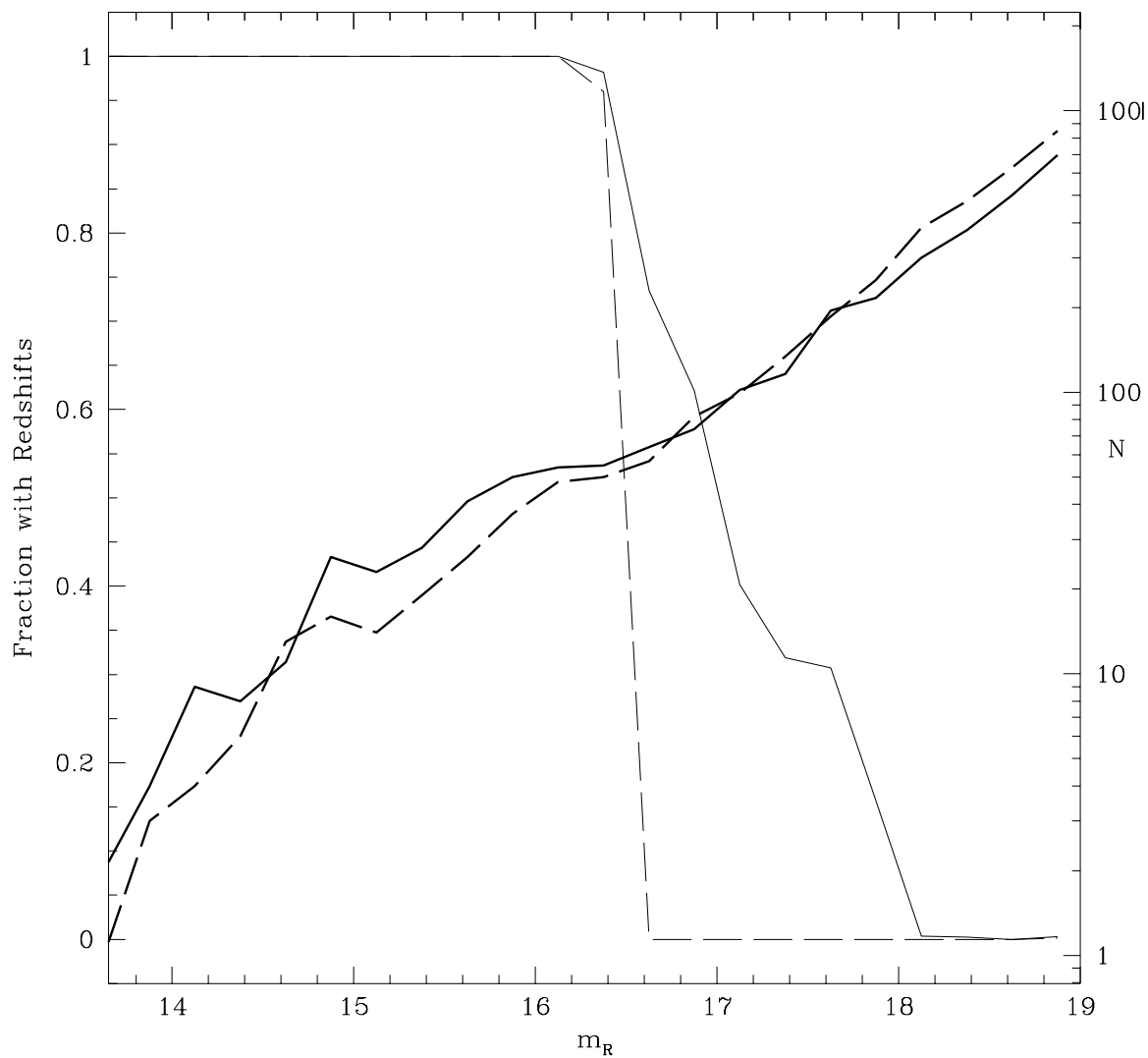} 
\caption{Fraction of galaxies with measured redshifts as a
function of $m_R$ magnitude. The solid line shows galaxies with
$R_p<1^\circ$, the dashed line shows galaxies at larger radii. We
also display the total number of galaxies in each bin (thick lines).}
\end{figure}

Two of us (JJM and GW) obtained 528 redshifts of galaxies in the
central $2^\circ \times 2^\circ$ of this region for a separate Jeans'
analysis of the central region of A576 (\cite{mw00}).
281 of these redshifts are from M96; the remainder will be published
in Mohr et al.~(2000). These redshifts were measured at the 
Decaspec (\cite{decaspec}) at the 2.4-m MDM telescope on Kitt Peak
and Hydra, the multifiber spectrograph on the WIYN telescope on Kitt
Peak. Figure \ref{figfracobserved} shows
the fraction of galaxies with redshifts as a function of R-band
magnitude. We divide the sample into the region mostly covered by JJM
and GW (projected radius ${R_p < 1^\circ}$) and the region mostly
covered by this study (${R_p > 1^\circ}$). Within the $3^\circ \times
3^\circ$ region covered by our images, the total sample of 817 galaxies
is 100\% (99.7\%) complete to a limiting magnitude of $m_R=16.2 (16.5)$;
the latter is the completeness limit for the photometric region.

\section{Galaxy Infall Method and the Mass Profile}

We briefly review the method DG and D99 develop to estimate the mass
profile of a galaxy cluster by identifying its caustics in redshift
space. The method assumes that clusters form through hierarchical
clustering and requires only galaxy redshifts and positions on the
sky. The amplitude $\mathcal{A} \mathnormal(r)$ of the
caustics is half of the distance between the upper and lower caustics in 
redshift space. Assuming spherical symmetry, $\mathcal{A} \mathnormal(r)$ is
related to the cluster gravitational potential $\phi (r)$ by 
\begin{equation}
\mathcal{A} \mathnormal^2(r) = -2 \phi (r)\frac{1-\beta (r)}{3-2\beta (r)}
\end{equation}
where $\beta (r)$ is the velocity
anisotropy parameter. DG show that the mass of a spherical shell of radii 
[$r_0,r$] within the infall region is given by the integral of the square of 
the amplitude $\mathcal{A} \mathnormal(r)$ 
\begin{equation}
\label{mprofeqn}
GM(<r)-GM(<r_0) = F_\beta \int_{r_0}^r \mathcal{A} \mathnormal^2(x)dx
\end{equation}
where $F_\beta \approx 0.5$ is a filling factor with a numerical value
estimated from simulations. Variations in
$F_\beta$ lead to some systematic uncertainty in the derived
mass profile. Our
mass profile extends to $\sim 3 r_{200}$ (the average density within
$r_\delta$ is $\delta$ times the critical density, $\S 4.1$);
within this 
radius, $F_\beta$ varies by $\lesssim 15\%$ in simulations (see
D99 for a more detailed discussion). 

Operationally, we identify the caustics as curves which delineate a
significant drop in the phase space density of galaxies in the
projected radius-redshift diagram. Galaxies outside the caustics are
also outside the turnaround radius. For a spherically symmetric
system, taking an azimuthal average amplifies the signal
of the caustics in redshift space and smooths over small-scale
substructures. D99 described this method in detail and showed that,
when applied to simulated clusters with galaxies modelled with
semi-analytic techniques, it 
recovers the actual mass profiles within a factor of 2 to
5-10$~\Mpc$ from the cluster center. D99 describes  some potential
systematic effects including projection effects and variation in the
galaxy orbit distribution $\beta (R_p)$. 

Simulations are necessary to
estimate the uncertainties due to projection effects and deviations
from spherical symmetry. In the simulations of D99, the degree of
definition of the 
caustics depends on the underlying cosmology; in simulations,
caustics are better defined in a low-density universe than a closed,
matter-dominated universe (D99). Surprisingly, the caustics of Coma,
A576, and 
several other clusters (\cite{rines2kb}) are generally better defined
than those of the simulated clusters. Thus, the uncertainties
estimated from these simulations might be overestimates. 

We apply the technique of D99 to our A576 survey to determine the spatial
and velocity center of the region as well as the location of the
caustics. In this technique, we determine the center of the system from
a hierarchical cluster analysis. The center thus derived, {$\alpha _{opt}
=$ 7:21:31.96}, {$\delta _{opt} = $+55:45:20.6} (J2000), $cz=11,487~\kms$,
lies 50$^{\prime \prime}$ ($28~\kpc$) from the X-ray center (M96, from
{\em Einstein} IPC data; see also $\S$ 4 for {\em ASCA} data). 
Our measurement of the central velocity of the infall region
agrees well with previous estimates (Struble \& Rood 1991, M96). 

\begin{figure}
\figurenum{4}
\label{figcaustics}
\plotone{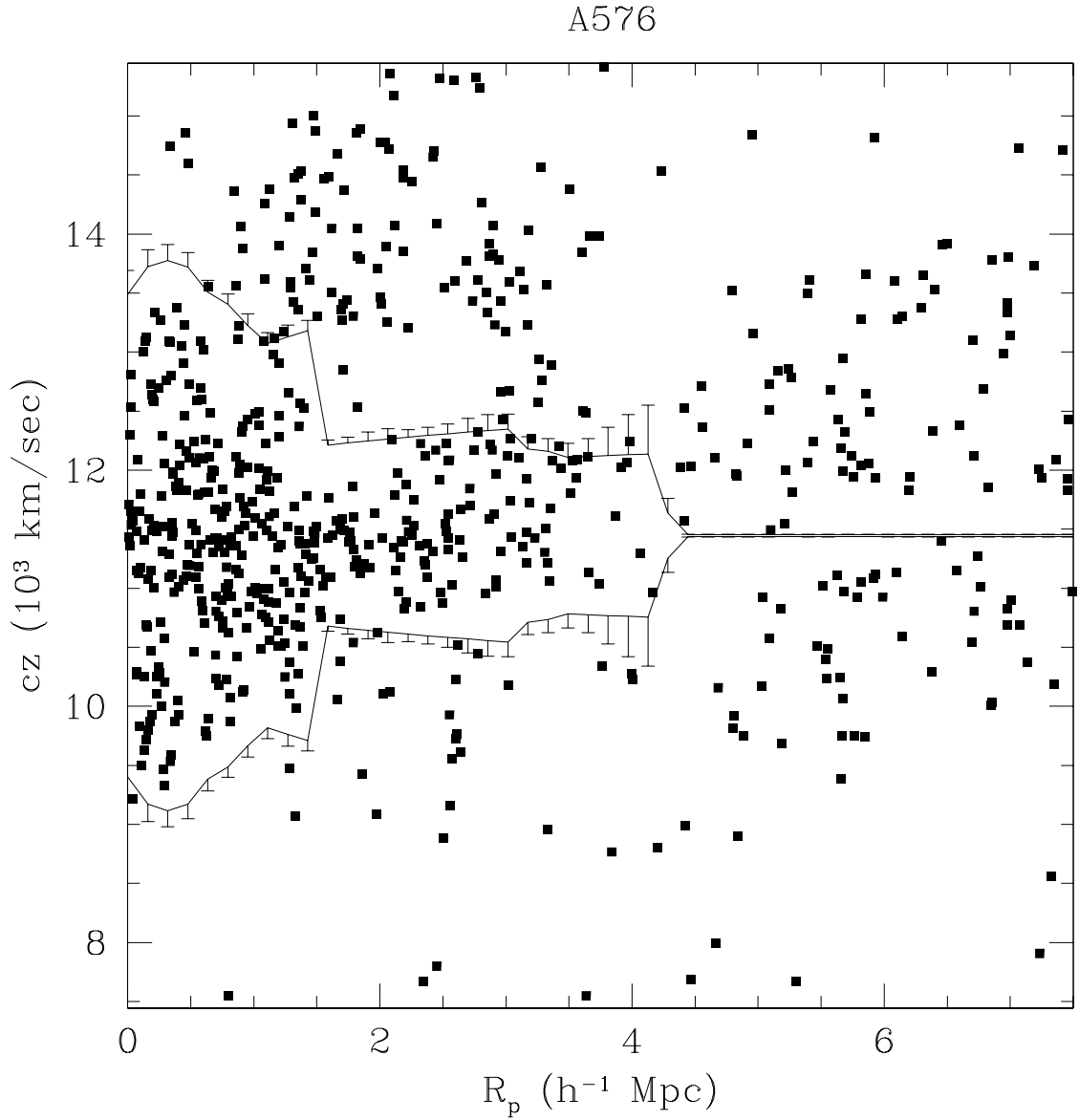} 
\caption{Redshifts as a function of projected radius in A576. The
solid lines are the caustics determined from our adaptive kernel
estimate with 1-$\sigma$ error bars.} 
\end{figure}

Figure \ref{figcaustics} shows the projected radius from the cluster
center versus redshift for galaxies within
$cz_{lim}=4000~\kms$ of the 
cluster center. This range of redshifts includes all galaxies which
may be members of the infall region. 
Solid lines indicate the caustics determined using
the method of D99 based on a multidimensional adaptive kernel
method (\cite{s86}; \cite{p93}; \cite{p96}). 
This technique has been applied to numerical simulations (D99) as well
as to the Coma cluster (GDK). As discussed in D99, it is necessary to
rescale $h_v$ and $h_r$, the velocity and radial smoothing lengths so
that spherical smoothing windows can be used. With an appropriate
choice of this scaling relation $q = h_v/h_r$, the location of
the caustics should be insensitive to small changes in $q$. In our
case, $q=25$ satisfies this criterion. Following D99, we define the
caustic amplitude $\mathcal{A} \mathnormal(r)$ as the 
minimum of the upper and lower amplitude 
estimates. This prescription is identical to averaging the two
estimates for an isolated, spherically symmetric system; taking the
minimum reduces the sensitivity to massive substructure and
contamination.

\begin{figure}
\figurenum{5}
\label{figcausticsq}
\plotone{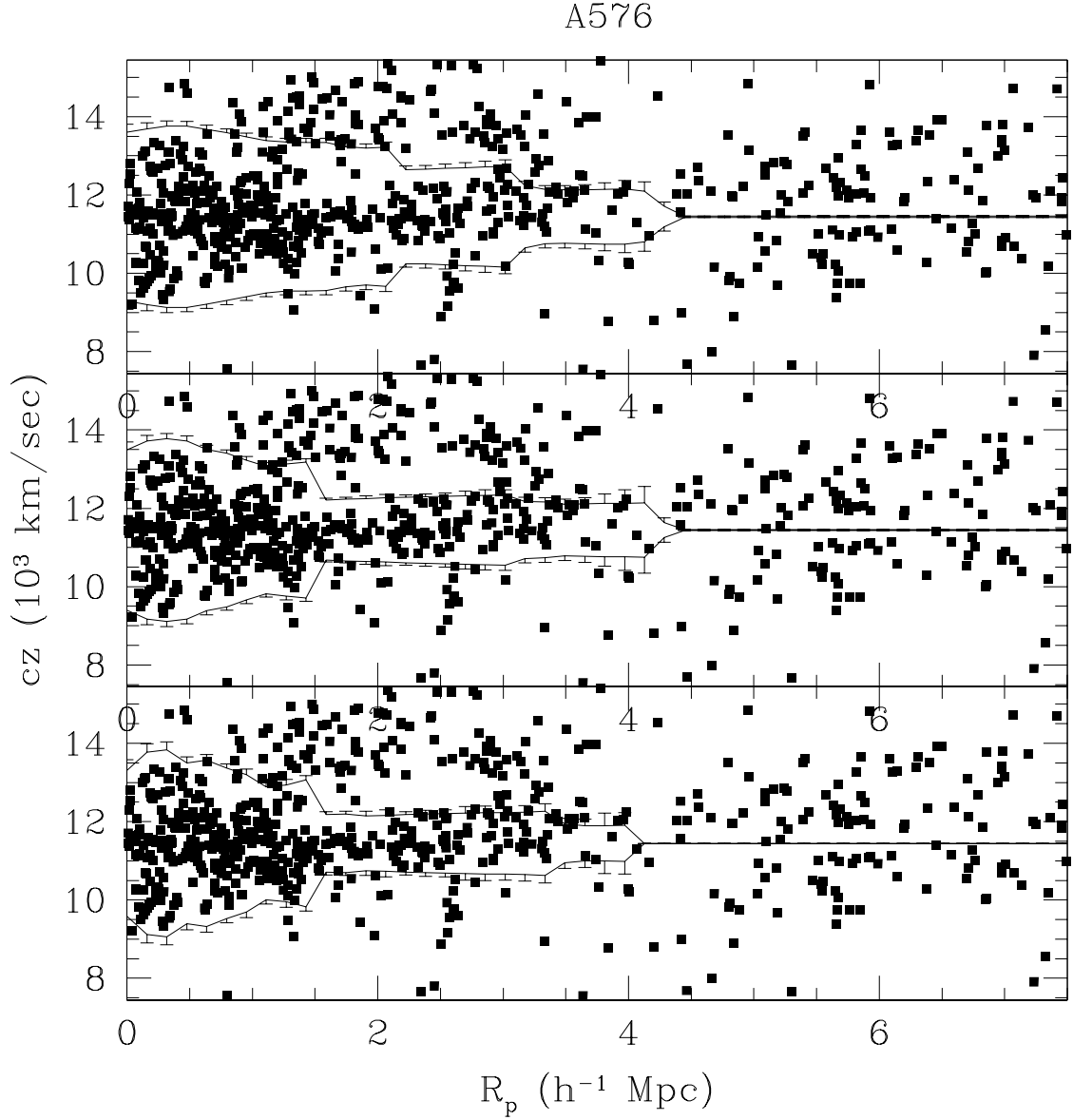} 
\caption{Dependence of caustic location on the parameter
$q$, the ratio of velocity uncertainty to positional uncertainty. From
top to bottom, the caustics are fit with $q=10$, 25, and 50.}
\end{figure}

We note that in A576, the location of the caustics is sensitive to
substructure at $1.5-2.2 \Mpc$. The caustic amplitude 
decreases sharply to $\sim 800~\kms$ somewhere in this region, though
the radius of the decrease is sensitive to the smoothing parameter
$q$. Figure \ref{figcausticsq} shows
the caustics determined by setting $q=10$, 25, and 50. Small values of
$q\sim 10$ seem to oversmooth the 
caustics and the sharp decrease occurs at $\sim 2.2 \Mpc$, whereas the
sharp decrease occurs at $\sim 1.5 \Mpc$ for
larger values of $q=25-50$. The amplitude of the caustics is stable
both at radii
smaller than $\sim 1.5 \Mpc$ and at radii larger than $\sim 2.2 \Mpc$. 
D99 gives the prescription for estimating
the uncertainties in the caustic amplitude; this prescription reflects  
the scatter due to projection effects in the simulations. We show
these $1\sigma$
uncertainties in Figure \ref{figcaustics}. 

\begin{figure}
\figurenum{6}
\label{figskyplotz}
\plotone{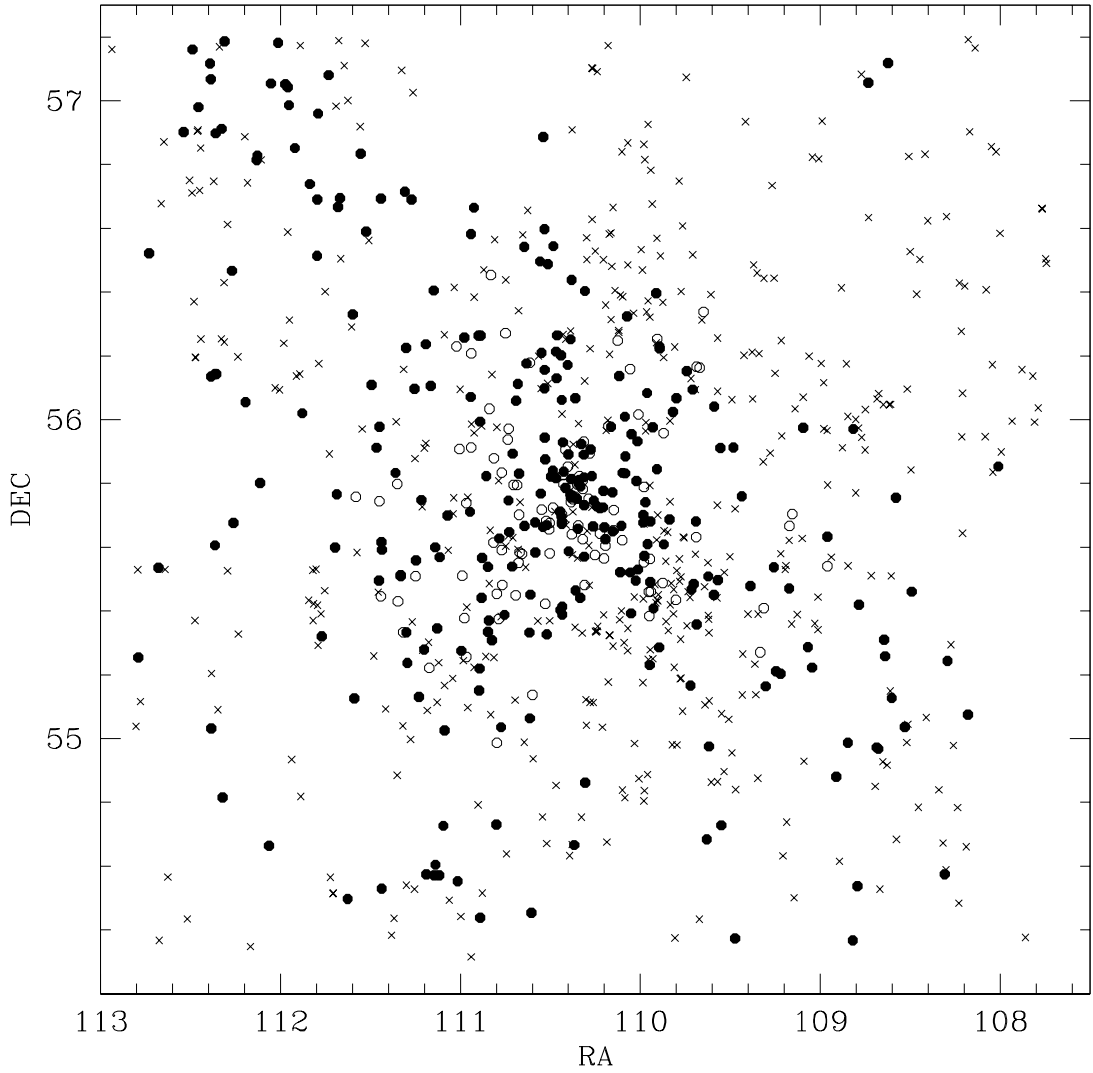} 
\caption{Distribution of cluster and field galaxies on the sky. Open
circles, filled circles, and crosses represent cluster members,
members with $m_R<16.5$, and background galaxies respectively.}
\end{figure}

We define membership
of the infall region from the caustics; hereafter, galaxies outside these
caustics are interlopers. Of the 497 galaxies within $cz_{lim}$ of the
velocity center of A576, 368 are within the infall region. Figure
\ref{figskyplotz} shows 
the distribution of these galaxies on the sky. There is a
noticeable deficit of infalling galaxies NW of the cluster center.

\subsection{Comparison of Mass Profile to Models}

We compare our results to  Navarro, Frenk, \&
White (1997, hereafter \cite{nfw97}),
Hernquist (1990), and  singular isothermal sphere mass profiles. The
Hernquist profile is an analytic form proposed as a model of elliptical
galaxies and bulges
(Hernquist 1990). The ``universal density profile'' of NFW
accurately models the mass profiles of dark matter halos in a variety
of cosmological simulations (\cite{nfw97}). Other simulations suggest
that the NFW profile may not be accurate at small radii (e.g.,
\cite{moore}), but these differences are unimportant on the scales
we probe here. These mass profiles are:   
\beqn
M_{NFW}(r) = C_{NFW} a [{\rm \ln}\Bigl(\frac{r+a}{a}\Bigr)-\frac{r}{r+a}],
\eeqn
\beqn
M_{Hern}(r) = C_{Hern} \frac{a r^2}{(r+a)^2},
\eeqn
\beqn
M_{iso}(r) = C_{iso} r
\eeqn
respectively, where $a$ is the characteristic radius and $C$ is a
normalization factor. These forms of the mass profiles
minimize the correlation between the parameters. For
the Hernquist profile, the mass $M_c$ within $a$ is $C_{Hern}
a/4$, and the total mass of the system is $M_{tot}=C_{Hern}a$.
From Equation \ref{mprofeqn}, a singular isothermal sphere
mass profile produces caustics with constant amplitudes. However,
Figure \ref{figcaustics} shows that the amplitude of 
the caustics decreases with radius. 

We fit these models to the observed mass profile by minimizing
$\chi^2$; Table \ref{mprof} lists the 
results. The measures of the cumulative mass profiles are not
independent; thus, the values of $\chi ^2$ are indicative and are only
meaningful when compared with each other. The best-fit parameters are
insensitive to the inclusion or 
exclusion of the three data points where the caustics are unstable
($1.5<R_p<2.2 \Mpc$). 
Figure \ref{figcmassprof} shows the three profiles which best fit the
infall mass profile and Figure \ref{figrho} shows the mass density
profile. The latter display
has the benefit that the data points are largely indepedendent of
one another. The density of the
cluster varies by five orders of magnitude across our sample; the
overall agreement between these profiles and the data is remarkable.  
The isothermal sphere profile is strongly excluded. The
Hernquist profile yields a 
better $\chi^2$ than the NFW profile, but both are
acceptable. This conclusion agrees with GDK, who found that the
mass profile of Coma is much better described by an NFW profile than
an isothermal sphere.

From the infall mass profile, we can 
derive the values of overdensity radii $r_\delta$ directly. The
mass $M_\delta$ contained within the overdensity radius $r_\delta$
exceeds the critical density by 
a factor of $\delta$. Two of the most
commonly used radii are $r_{200}$ and $r_{500}$. The values of
these radii are usually determined indirectly. For instance, Carlberg
et al.~(1997, hereafter CYE)
define $r_{200}$ using the velocity dispersion $\sigma$ of 
a cluster. Evrard, Metzler, \& Navarro (1996) use simulations to
calibrate a correlation between $r_{500}$ and the emission-weighted
X-ray temperature $T_X$. 
We estimate $r_{500} = 0.96 \pm 0.05 \Mpc$ and
$r_{200} = 1.42 \pm 0.07 \Mpc$. A576 has an emission-weighted
temperature of $T_X = 3.77$ keV (see $\S 5$), which yields an estimate
of $r_{500} = 0.76 \pm 0.16\Mpc$ using the Evrard et al.~estimator, in
agreement with our result.

\subsection{Velocity Dispersion Profile}

Many recent papers analyze the velocity dispersion profiles of
clusters (e.g., \cite{f96}; \cite{dk96}; M96). When combined with the
galaxy number density profile in the Jeans equation, the velocity
dispersion profile can provide an estimate of the 
mass profile; Mohr et al.~(2000) will perform this analysis for
A576. Figure \ref{figvdp} shows
the velocity dispersion profile of A576 where we compute the
dispersions in bins of 25 
galaxies. We also display the cumulative 
projected velocity dispersion profile $\sigma _p(<R_p)$ (calculated from
all galaxies inside $R_p$).
The profile is centrally peaked and would probably be classified as
'Peaked' by den Hartog \& Katgert (1996).  

Several authors (\cite{f96}; CYE) suggest
that the velocity dispersion of a cluster is best estimated by the
asymptotic value of the cumulative projected velocity dispersion. For
A576, $\sigma _p(<R_p)$ decreases monotonically with radius for $R_p
\gtrsim 0.5 \Mpc$, suggesting that an asymptotic
value of $\sigma _p(<R_p)$ may not exist. At the largest radius we
study,  $\sigma _p(<R_p) \approx 800~\kms$. This velocity dispersion
is significantly smaller than $\sigma 
_p = 977^{+124}_{-96}~\kms$ (M96) or $\sigma
_p = 914^{+50}_{-38}~\kms$ (Girardi et al.~1998). 

The CYE estimate of $r_{200}$ is sensitive to the definition
of $\sigma _p$.
We obtain $r_{200} = 1.69 \pm 0.17 \Mpc$
using the M96 value of $\sigma _p$, 1.5$\sigma$ larger than our
estimate of $r_{200} = 1.42 \pm 0.07 \Mpc$ ($\S 3.1$). Taking the
value of $\sigma _p(<R_p) \approx 800~\kms$ from the limit of our
survey, $r_{200} = 1.36 
\pm 0.15 \Mpc$, in agreement with our infall estimate. The sensitivity
of the CYE estimator to the
aperture used to measure $\sigma _p$ may affect many of the
results from studies of CNOC clusters because the properties of the
composite CNOC cluster 
depend on the estimates of $r_{200}$ for individual
clusters. 

Based on the infall mass profile, we can predict the velocity
dispersion profile with an assumption about the distribution of galaxy
orbits. We assume that orbits are isotropic at all radii ($\beta =
0$), and calculate the velocity dispersion profiles for the Hernquist
and NFW models (\cite{h90}, Equation 10; NFW, Equation 3). These
predicted velocity dispersion profiles agree with the observed
profile (Figure \ref{figvdp}). The largest differences are at the
radii where the caustic amplitude is unstable. Without fitting any
parameters, we find $\chi ^2 = 42$ and 59 for 14 degrees of freedom
for the Hernquist and NFW profiles respectively. This result provides a
consistency check on the infall mass profile and confirms that 
a Hernquist profile models the data better than an NFW profile. 

\subsection{Alternative Kinematic Mass Estimators}

The two most commonly used kinematic mass estimators for 
clusters are the virial mass estimator and the projected mass
estimator (\cite{htb85}). These estimators both assume that 
clusters are relaxed systems. Numerical simulations, however, suggest
that both of these mass estimators typically overestimate the true
mass profile of a 
relaxed cluster by $\sim 20\%$ (\cite{ap99}). 

\begin{figure} 
\figurenum{7}
\label{figcmassprof}
\plotone{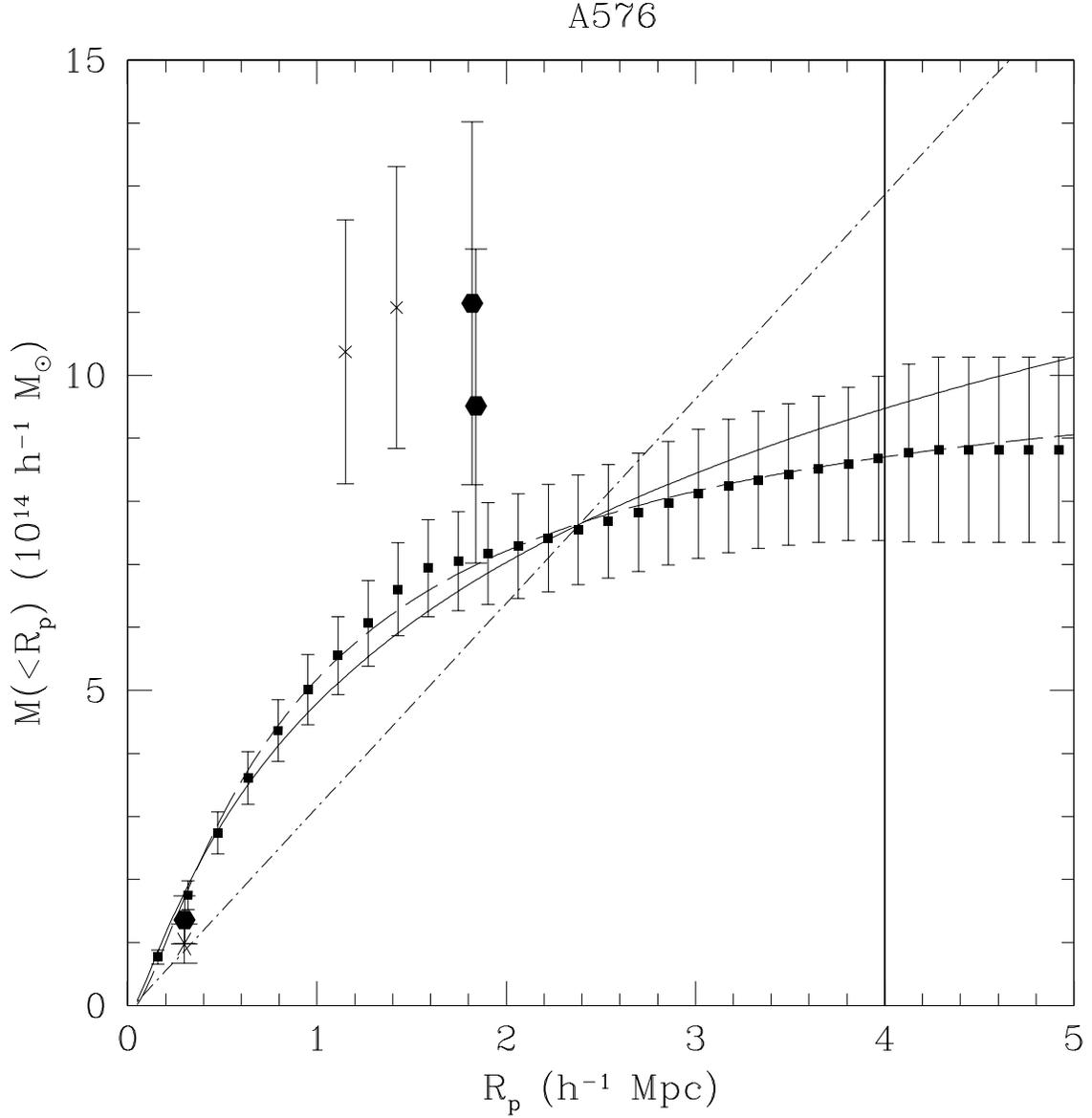} 
\caption{Mass profile of A576. The filled squares are from
the adaptive kernel estimate of the caustics with $1-\sigma$
uncertainties shown. The upper and lower
filled hexagons are virial mass estimates (Girardi et
al.~1998) omitting and including the surface term
respectively; the filled hexagon at small radius is an interpolation
by Girardi et al.~discussed in the
text. Crosses show our virial and projected mass 
estimates, the star shows an X-ray mass estimate ($\S 4.1$, \cite{wjf97}). The
solid, dashed, and dash-dot lines are the best-fit NFW, 
Hernquist, and isothermal sphere profiles respectively.}
\end{figure}

\begin{figure} 
\figurenum{8}
\label{figrho}
\plotone{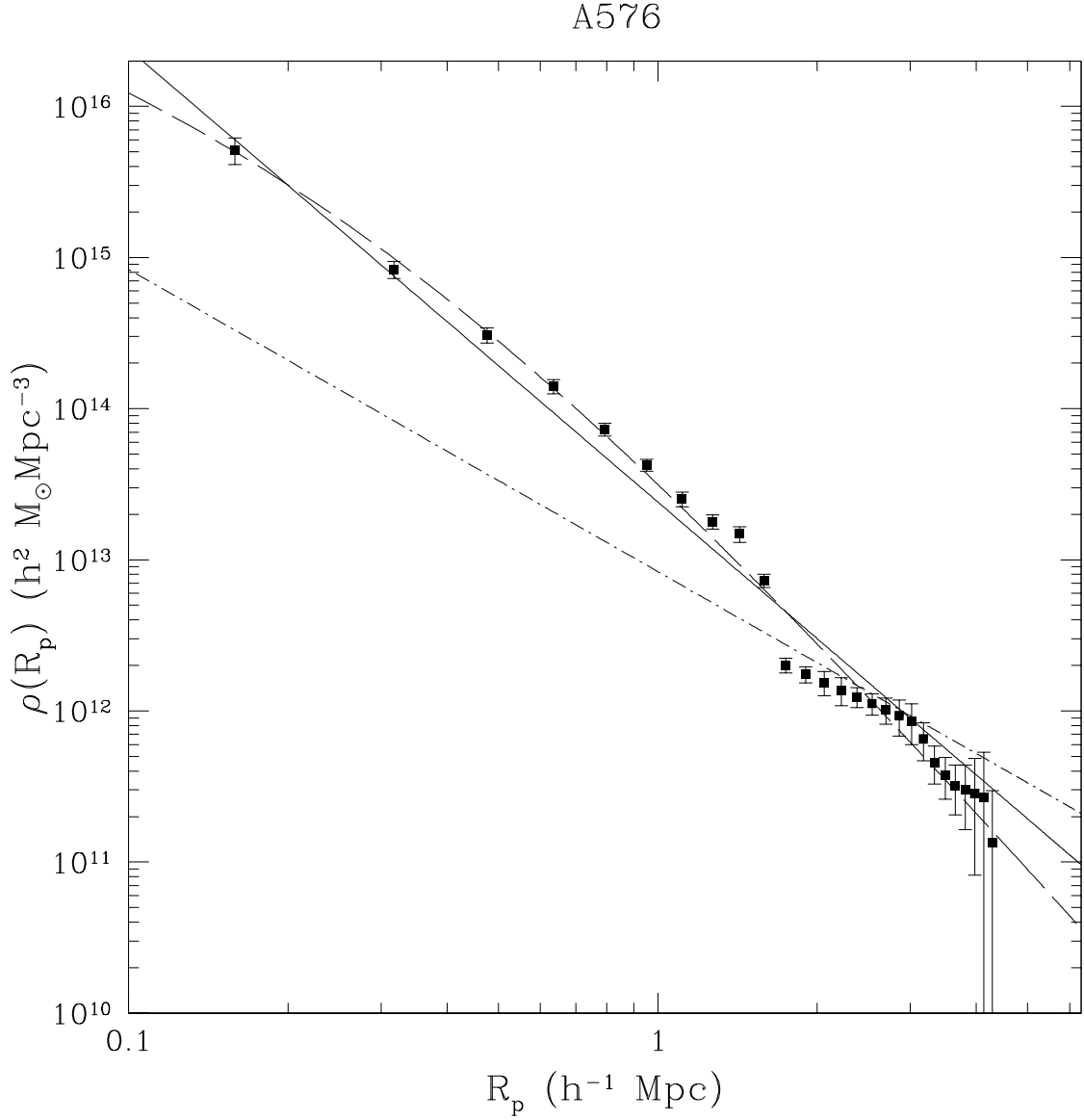} 
\caption{Mass density profile of A576. Lines have the same definitions
as in Figure \ref{figcmassprof}.}
\end{figure}

\begin{figure} 
\figurenum{9}
\label{figvdp}
\plotone{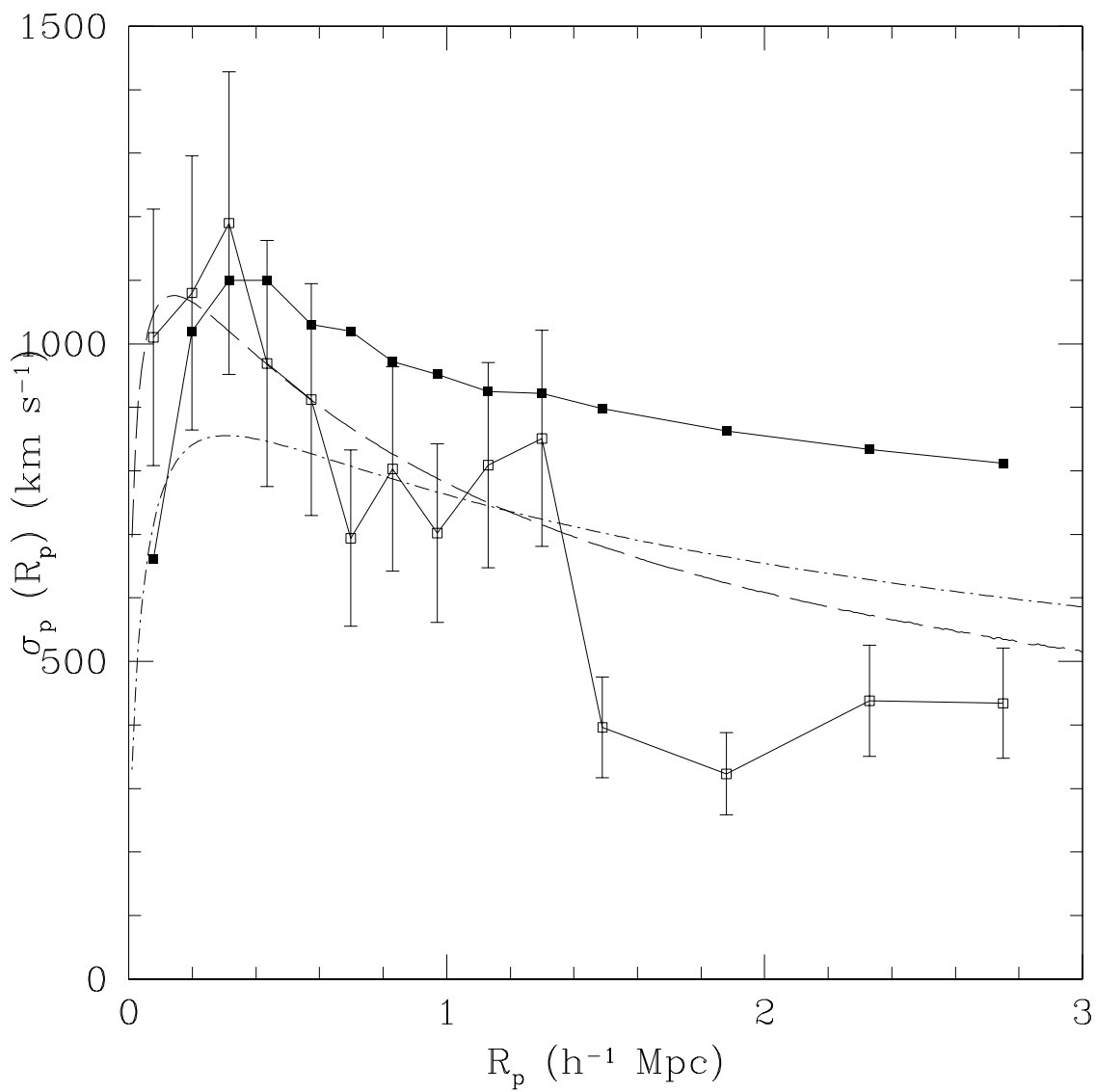} 
\caption{Velocity dispersion profile for A576. Open squares show the
velocity dispersion profile $\sigma _p(R_p)$. Filled squares show the
cumulative velocity dispersion profile $\sigma _p(<R_p)$. Dashed and
dash-dotted lines indicate the predicted profiles for the best-fit
Hernquist and NFW profiles respectively assuming isotropic orbits
($\beta = 0$).} 
\end{figure}

If clusters are not relaxed, these estimators may provide more
substantial overestimates of the mass.
Figure \ref{figcmassprof}  shows the virial mass estimates of A576
calculated by Girardi et al.~(1998) from the M96 data. They report
both the standard virial mass and a corrected virial mass which
includes a correction for the surface pressure term  estimated from
the galaxy distribution. This correction assumes that light traces
mass, or more precisely, that the number density of galaxies traces
mass.  Girardi et al.~also use the galaxy number density profile to
estimate the mass of A576 at small radius for comparison with X-ray
estimates.

M96 show that their data allow a wide range of
masses ($0.6\rightarrow 1.5 \times 10^{15} \msun$) depending on the
mass estimator, the magnitude cutoff, and the definition of cluster
members. In particular, emission dominated galaxies have a larger
velocity dispersion than absorption dominated galaxies. The virial
mass of emission line galaxies is a factor of $\sim 2$ larger than the
virial mass of absorption line galaxies (M96; \cite{c97a}). Note,
however, that when we combine the galaxy number density and the
velocity dispersion profiles in the Jeans equation, the two subsamples
should yield consistent mass profiles (e.g., \cite{c97a}). M96
conclude that their data are  
insufficient to constrain the mass of A576 well. The M96 mass range 
encloses the mass estimates and uncertainties given by Girardi et
al.~(which make no correction for galaxy populations) as well as our
infall mass estimate. 

Restricting our analysis to galaxies within $r_{200} = 1.42 \Mpc$ ($\S
3.1$), we use the projected mass estimator to estimate 
$M = (11.1\pm 2.2) \times 10^{14}~\msun$ within $r_{200}$ assuming
isotropic orbits; applying the virial theorem yields $M = (10.4\pm 2.0)
\times 10^{14}~\msun$ within $r_{vir} = 1.15 \pm
0.2 \Mpc$. We display all of these estimates in Figure
\ref{figcmassprof}.  All of the mass estimates at large radius exceed
the infall estimate, although the corrected estimate by Girardi et
al.~is consistent with the infall mass profile. 

\section{X-ray Data and Analysis}

A576 has been observed by both {\em Einstein} and {\em ASCA}. $ASCA$'s
broad energy band (0.5-10.0 keV) is particularly useful for  
determining cluster temperatures. Because of the poor angular
resolution of $ASCA$, we determine the emission weighted average
temperature within $15'$, or $\sim 0.5~\Mpc$. We obtained 
the screened data from GSFC. We extract a spectrum including all photons
within a circle of radius $15'$ (60 pixels) 
centered on the cluster center for GIS data from a long (97 ksec)
observation. The centroid of the SIS 
image agrees with the position of the {\em Einstein} centroid within
the $\sim 0\farcm 4$ uncertainty in the SIS position (\cite{ascaposn}).

Using XSPEC (v10.0), we fit the cluster spectrum to a model including
absorption (`wabs') parameterized by the column density
of hydrogen (which we set to the galactic value) 
and the standard Raymond-Smith model (\cite{rs77}) characterized by
temperature, iron abundance, redshift, and a normalization factor. The
iron abundance is measured relative to cosmic abundance. 
We fit the temperature, iron abundance, and normalization as free
parameters. 

We use the weighting system
developed by Churazov et al.~(1996), which avoids
rebinning the data into broad bins. Because there are few counts above
8.0 keV, we only include data from  $0.8 \to 
8.0$ keV, though we fit the spectrum to  
slightly different ranges to ensure that the fitted model parameters
are consistent for different choices. 

\begin{figure} 
\figurenum{10}
\epsscale{0.75}
\label{figxrayspectrum}
%\plotfiddle{f10.eps}{600}{-90}{100}{100}{0}{0}
\plotone{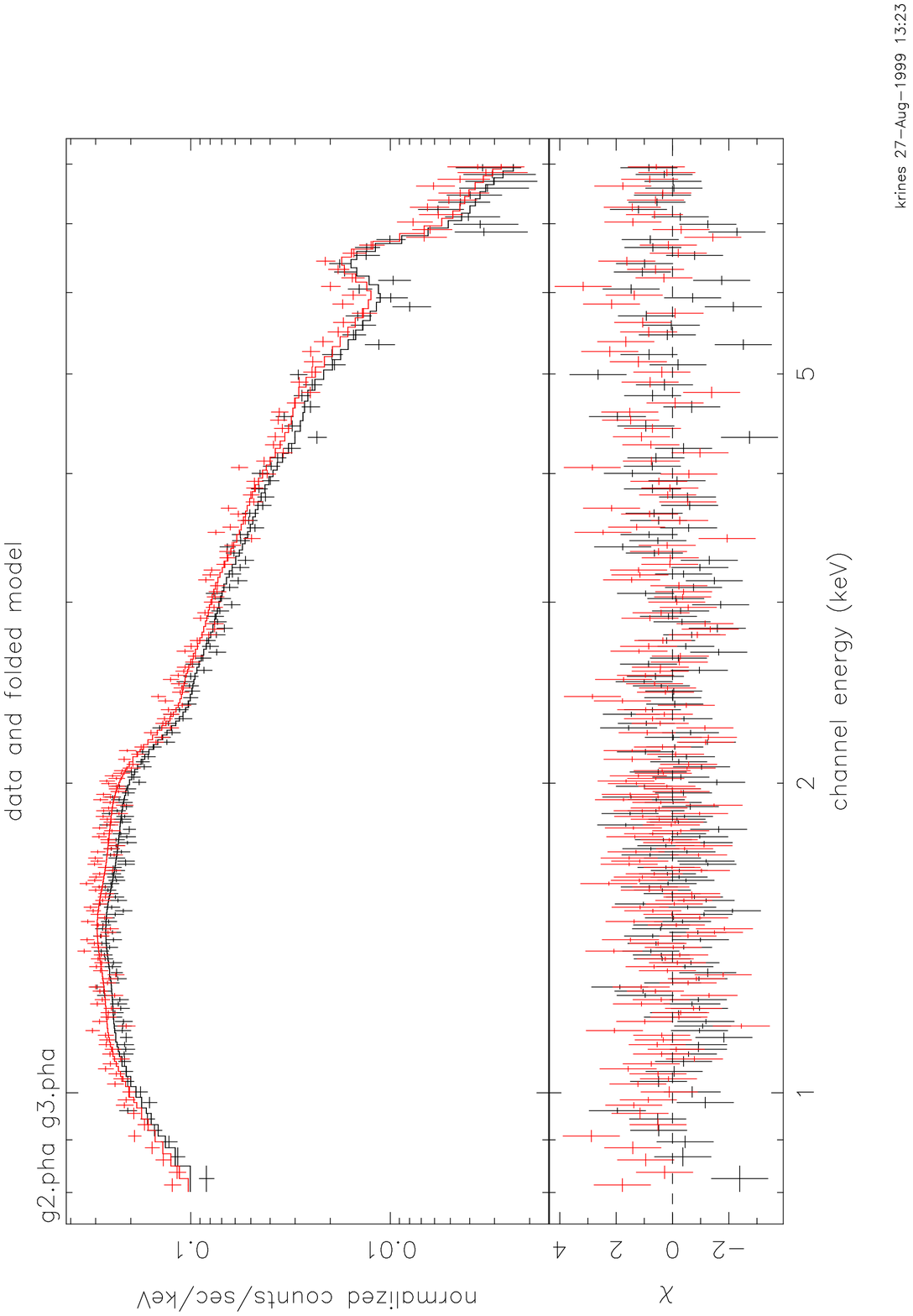}
\caption{{\it ASCA} GIS spectrum of A576. The solid line is the best-fit
single-temperature Raymond-Smith thermal plasma model.}
\end{figure}

We obtain acceptable fits by assuming that the gas is isothermal. 
More complicated models are thus unnecessary. Our best-fit model has an ICM
temperature of 3.77$\pm 0.10$ keV with an iron abundance
0.27$\pm 0.05$ cosmic. Uncertainties are 68\%
confidence limits for one 
parameter.  Figure \ref{figxrayspectrum} shows the X-ray spectrum and
the best-fit model. This temperature is 1.7$\sigma$ less than
the temperature of 4.3$\pm 0.3$ keV (David et al.~1993) from
{\em Einstein} MPC data and 2.5$\sigma$ less than $4.02\pm0.07$ keV
from an independent analysis of the {\em ASCA} data (\cite{wtp00}). 

We calculate the expected flux between 0.01 and
100.0 keV from the best-fit model, yielding an essentially bolometric
flux of $4.30 \times 10^{-11}$ ergs cm$^{-2}$ s$^{-1}$.
We calculate a bolometric luminosity of $0.71 \times 10^{44}
h^{-2}$ ergs s$^{-1}$ 
(we assume $q_0$ = 0.0) in agreement with $0.73 \times 10^{44}
h^{-2}$ ergs s$^{-1}$ from {\em Einstein} MPC data (\cite{david93}). 
The 0.3-3.5 keV luminosity is $0.46 
\times 10^{44} h^{-2}$ergs s$^{-1}$, in excellent agreement with
M96, who find a luminosity of $0.45 
\times 10^{44} h^{-2}$ergs s$^{-1}$ from the {\em Einstein} IPC data
assuming an ICM temperature of \hbox{4.3 keV}. 

M96 found  
that a two-temperature Raymond-Smith model improved the fits to the
{\em Einstein} spectrum. We test this result by adding a second
thermal component 
to our isothermal model, but the fitting routine forces the
temperature of the second component to be extremely small ($\lesssim
0.1$keV), suggesting that no second component is needed to explain the
more complete {\em ASCA} data. This result contradicts the lower central
temperature found with {\em Einstein} SSS data (\cite{roth}). An
independent anaysis of the {\em ASCA} data (\cite{wtp00}) suggests that
A576 has a flat temperature profile  and
no cooling flow. 

We use {\em Einstein} IPC data to analyze the surface brightness
profile. Previous studies (\cite{jf84}; 1999; M96) fit the profile to
the hydrostatic-isothermal $\beta _x$ model (\cite{cff76}) and 
found $\beta _x = 0.45-0.53$, and 0.58-0.72 and
core radius $a = 50-70,$ and $90-160~\kpc$ for Jones \& Forman
(1999) and M96 respectively. These models yield central gas and
electron number densities
of $\rho_0 = 3.6 (2.3) \times 10^{-27}$ g cm$^{-3}$ and $1.9 (1.2)
\times 10^{-3} $cm$^{-3}$ respectively. M96 
suggest that the disagreement arises from fitting
different radial ranges; Jones \& Forman extend their analysis to
larger radius than M96. We consider both models below.

\subsection{X-ray Mass Profile}

With the assumptions of hydrostatic equilibrium, negligible
non-thermal pressure, and spherical symmetry,
the gravitational mass inside a radius r is 
\begin{equation}
\label{mass1}
M_{tot}(<r) = - \frac{kT}{\mu m_p G} \Bigl({{d\ln \rho_{gas}} \over
{d\ln r}} + {{d\ln T} \over {d\ln r}} \Bigr) r
\end{equation}
 (\cite{flg80}). For a uniform temperature distribution, the second
term on the right hand side vanishes. We then only need to
determine the gas temperature and the density distribution of the gas
to calculate the gravitational mass. 
Under the standard hydrostatic-isothermal $\beta _x$ model, 
the mass is related to $\beta _x$ and $a$ by 
\begin{equation}
\label{mass2}
M_{tot}(<r) = \frac{3kT\beta _x r^3}{\mu m_p Ga^2(1 + (\case{r}{a})^2)}
\newline = 5.65\times 10^{13} \beta _x T_{\mbox{keV}} \frac{r^3}{a^2 +
 r^2} \msun
\end{equation}
where $M_{tot}(<r)$ is the total gravitational mass within a radius
$r$ and the numerical approximation is valid for $T_{\mbox{keV}}$ in
keV and $r$ and $a$ in $\Mpc$. 

\begin{figure} 
\figurenum{11}
\label{figxraymassprof}
\plotone{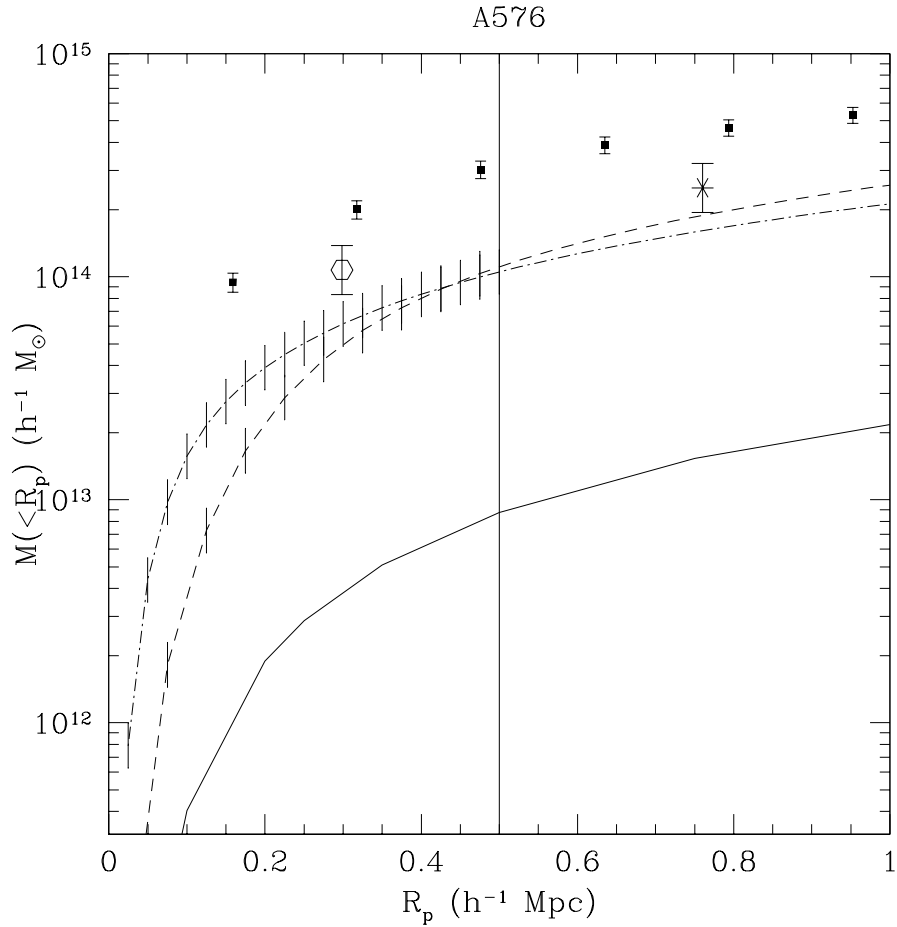} 
\caption{Mass profile of A576 derived from X-ray observations under
the assumption of the hydrostatic-isothermal $\beta _x$ model and
parameters from Jones \& Forman (1999, dash-dotted line) and
M96 (dashed line). The solid line is the gas mass profile and the
vertical line  at $R_P = 0.5~\Mpc$ shows the limit of the X-ray
data. Filled squares show 
the infall mass profile. The open hexagon and star are X-ray mass estimates
from White et al.~(1997) and from the estimator of Evrard et
al.~(1996) respectively.}
\end{figure}

We use Equation~\ref{mass1} to calculate the total mass profile of
A576 within 1.5$\Mpc$; the estimate outside $\sim 0.5~\Mpc$ is an
extrapolation. We use our {\em ASCA}-derived temperature with the results
of both M96 and Jones \& Forman (1999) to estimate the mass
profile. Figure \ref{figxraymassprof} shows these profiles as well as
the gas mass profile for the M96 parameters. We also display an
estimate of the cluster mass based on spherical deprojection (White,
Jones \& Forman 1997). This estimate is $M \approx 1.07 \times
10^{14}~\msun$ at $r=0.299~\Mpc$ and is larger than either $\beta _x$
model estimate. 

Evrard et al.~(1996) develop a method to
reduce the scatter in cluster mass estimates by relying solely on the
emission-weighted gas temperature $T_X$ within a radius where the mean
density is 500 times the critical density. This radius, denoted by
$r_{500}$, varies with $T_X$ as 
\begin{equation}
r_{500} = (1.24 \pm 0.09) \Bigl( \frac{T_X}{10  \mbox{keV}} \Bigr)^{1/2}
h^{-1} \mbox{Mpc} .
\end{equation}
The mass within $r_{500}$ is approximately
\begin{equation}
\label{massev}
M_{500}(T_X) = 1.11 \times 10^{15} \Bigl(\frac{T_X}{10  \mbox{keV}} \Bigr)
^{3/2}~\msun
\end{equation}
and has an average estimated-to-true mass ratio of 1.00
with a standard deviation of 8-15\%. For A576, this procedure yields
$M_{500} = 2.57 \times 10^{14}~\msun$ at $r_{500} = 0.76 \pm 0.16~\Mpc$. This
semi-empirical estimate of $r_{500}$ agrees with our more
direct estimate of $r_{500} = 0.96 \pm 0.05~\Mpc$ from the
caustics. From the infall mass profile, we estimate $M_{500} = (5.1\pm0.5)
\times 10^{14}~\msun$.

Within $\sim 0.5~\Mpc$, the infall mass profile is a factor of
$\sim$2.5  larger than the 
isothermal X-ray mass profiles and a factor of $\sim$1.8 larger than
the deprojection estimate of White et al.~(1997). M96 find a similar
difference between X-ray mass estimates and other kinematic mass
estimates of A576. These differences could be due to non-thermal
pressure support, nonisothermality, or asymmetry.

\subsection{Gas Mass Fraction}

\begin{figure} 
\figurenum{12}
\label{figfgr}
\plotone{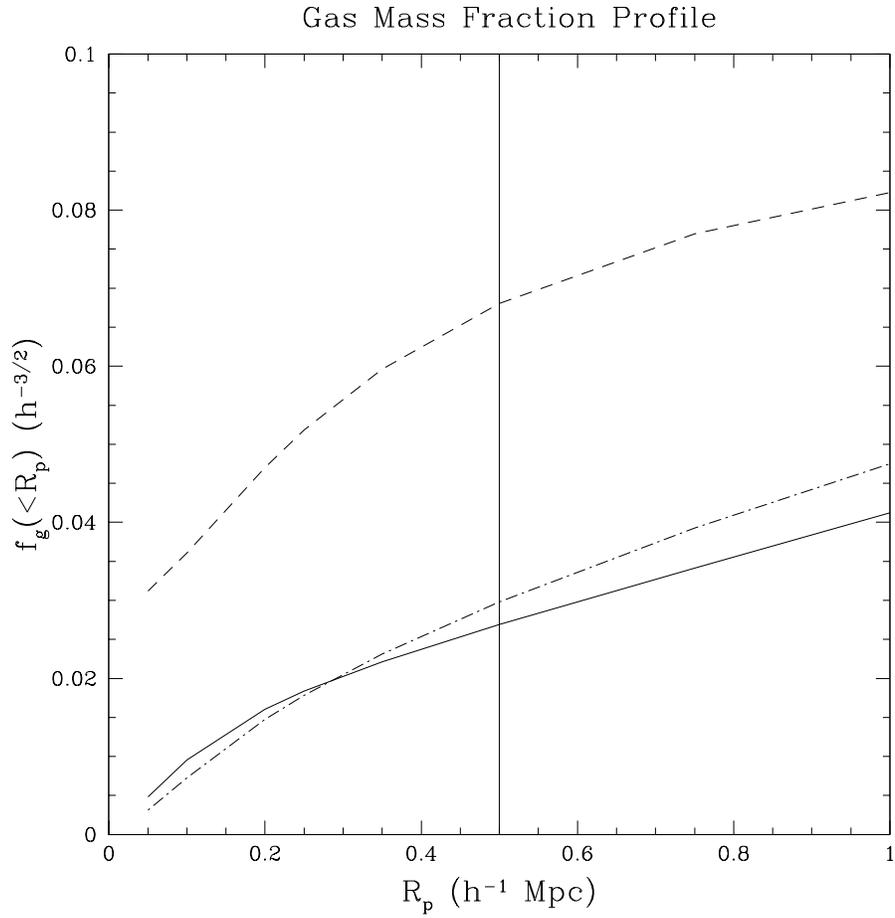} 
\caption{Gas mass fraction profile. $M_{gas}$ is taken from the X-ray
temperature and density profile, $M_{tot}$ is the best-fit X-ray mass
profile (dashed line) or Hernquist (solid) or NFW (dash-dot)
infall mass profile. The vertical line is the limit of the X-ray data.}
\end{figure}

Using only X-ray data, the gas mass fraction $f_g$ (the
fraction of total gravitational mass in hot gas) increases with radius
(Figure \ref{figfgr}), 
suggesting that the hot gas is more extended than the mass
distribution, in agreement with previous studies
(\cite{djf95}; \cite{mv97}; \cite{ef99}). The gas mass fraction is
$\sim 0.07 h^{-3/2}$ at $R_p = 0.5~\Mpc$,  less than (but consistent
with) the average value of gas mass fractions in the most luminous
clusters (e.g., \cite{white93}; \cite{e97}; \cite{mme};
\cite{ef99}). In hydrodynamical simulations (e.g. \cite{e97} and
references therein), the hot gas is more extended than the dark matter
because of the shock heating originating during the merging of the
halos which will form the cluster. In real systems, energy ejection from
supernovae could also contribute to heat the gas.

We observe a similar trend of increasing gas mass fraction with radius
when we use the best-fit NFW 
infall mass profile to calculate the gas mass fraction (Figure 
\ref{figfgr}). Because the infall mass profile exceeds the X-ray
mass profile, the inferred gas mass fraction ($\sim 0.03
h^{-3/2}$ at $R_p = 0.5~\Mpc$) is smaller.

\section{Estimating the Contribution of Background Galaxies}

Within the completeness limit of our redshift survey, the elimination
of background galaxies is straightforward; we simply remove
non-member galaxies using the caustics in the
projected radius-redshift diagram. We must estimate the background
statistically at magnitudes fainter 
than $m_R=16.5$. For $16.5<m_R<18.0$, we only have spectroscopy in the
central region ($< 0.^\circ 75$ radius; \cite{mw00}). In this central
region, the survey is 
91, 79, and 60\% complete to limiting magnitudes of $m_R=17.0$,
17.5, and 18.0 respectively. We assume that this region is sufficiently
large to represent a fair sample of the background over the entire
field. We then calculate the number of background
galaxies in each 
0.5 mag bin by assuming that the fraction of cluster members is the
same for galaxies with and without redshifts.

\begin{figure}
\figurenum{13}
\label{figpisanifv}
\plotone{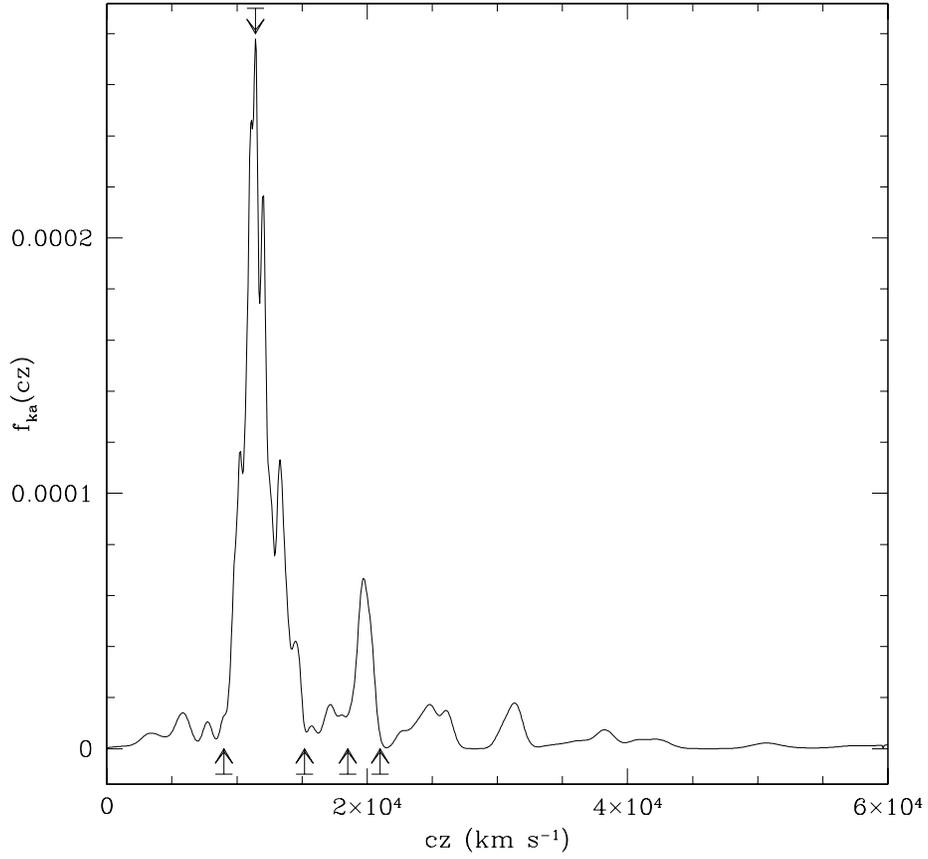} 
\caption{Adaptive kernel estimate of $f_{ka}(cz)$, the velocity distribution
function. Arrows indicate the peak and limits
of A576 and the limits of a background concentration of galaxies.}
\end{figure}

\begin{figure}
\figurenum{14}
\label{figbkgdstuff}
\plotone{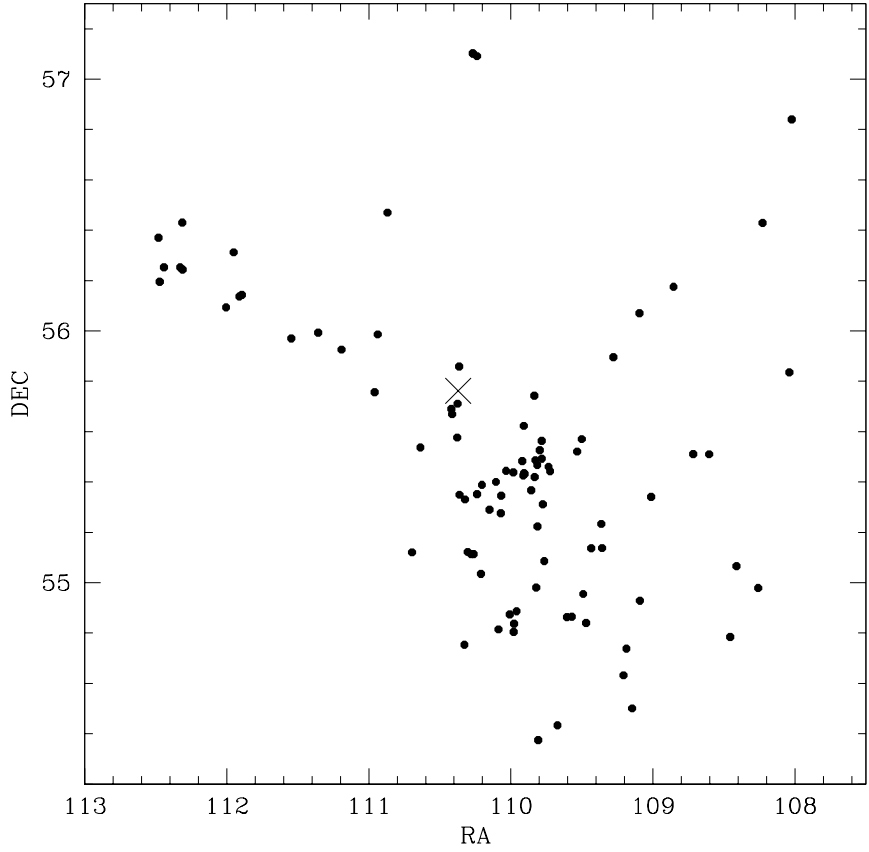} 
\caption{Distribution on the sky of background concentration of
galaxies. The cross marks the X-ray center of A576.}
\end{figure}

It is important to determine whether the survey of A576 contains any 
background or foreground groups or clusters of galaxies; these groups 
would lead to a 
localized enhancement in the number density of galaxies.
Applying the adaptive kernel method (\cite{p93}) to all redshifts
in our sample 
yields the estimated parent distribution shown in Figure
\ref{figpisanifv}. A576 is very prominent.
This distribution shows a background
concentration of galaxies at $19,700~\kms$; these galaxies
concentrate in a region $\sim 0.^\circ 50 - 0.^\circ 75$
from the center of A576 (Figure \ref{figbkgdstuff}). This
concentration suggests an enhancement of 
background light in this region relative to a randomly selected
field. Because we use this region to estimate the background, we may
overestimate the background luminosity density across the entire
region. However, the excess background  may be present across the
entire field.  

\begin{figure} 
\figurenum{15}
\label{figcscounts}
\plotone{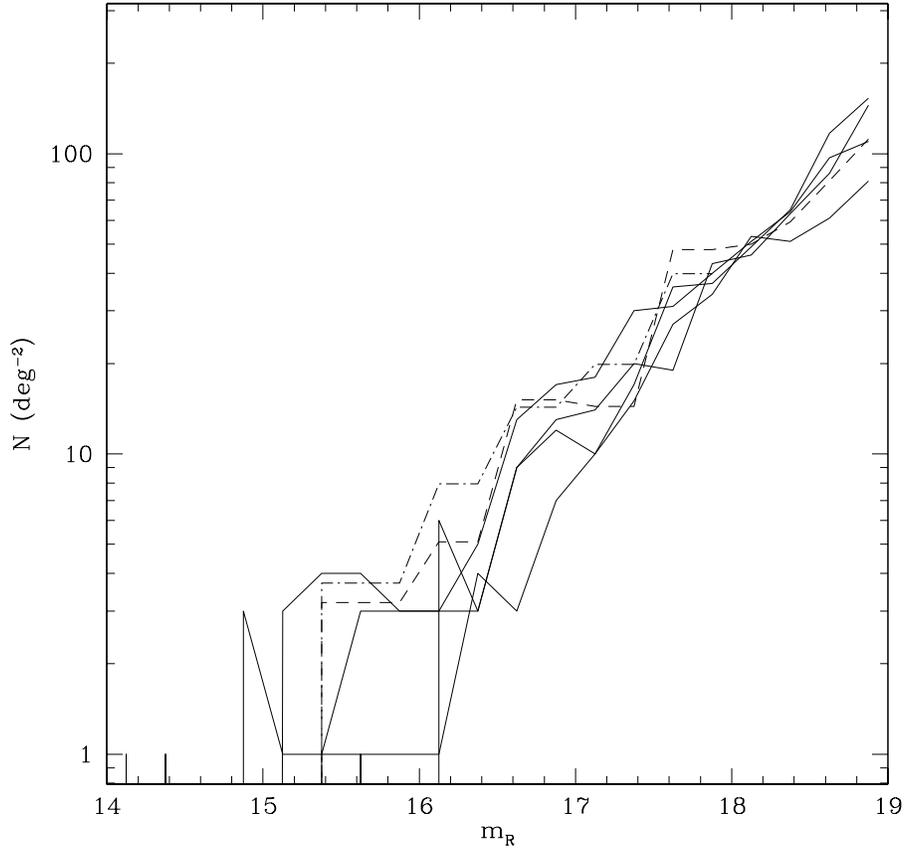} 
\caption{Apparent magnitude distribution of four randomly selected fields
from the Century Survey (solid lines). The dashed line is the
background estimated 
from the central region of A576; the dash-dot line is estimated from
the asymptotic number density profile.}    
\end{figure}

To estimate the variation of galaxy backgrounds
in randomly selected fields, we analyze four $1^\circ \times 1^\circ$
images of fields in the 
Century Survey (\cite{warren}) taken on the same nights with the
same observing setup as the A576 data. Due to large-scale structure,
the background counts vary more than expected from a
Poisson distribution; the mean number of galaxies in a field is 121
and the variance is 19.2 (the four fields contain 149, 118, 110, and
107 galaxies). Weighted by luminosity, the variance is
21\%. Figure \ref{figcscounts} shows the magnitude
distributions from these fields. For comparison, we show the estimate
from the central region of A576. In the magnitude range of interest
($16.5<m_R<18.0$), the estimate of background light from
the average of the Century Survey fields is $\sim 40\%$ (roughly 2
$\sigma$) smaller than that from the central region of A576. 
Because this enhanced background may or may not be present across the
entire region, we estimate that the
background subtraction is accurate to $\sim 40\%$ on scales
$\gtrsim 1^\circ$ square.

A third method of estimating the background is to assume that the
surface number density of cluster galaxies becomes negligible at the
outermost radii of our survey; we require that the cluster number density
be no less than that of known cluster galaxies. We attribute
additional surface number density  to background
galaxies. By measuring this number density as a function of limiting
magnitude, we can estimate the number counts of background galaxies
and the total background light.  Figure
\ref{figcscounts} shows the magnitude-number distribution from this
method. Because this method assumes that no faint galaxies at large
radii are infall members, this method should overestimate the 
background light.
The ratio of the background light estimates
calculated from the central region, the number density profile, and
Century Survey fields are 1.14 : 1 : 0.67.

The background estimates from the central region of A576
and from its outskirts agree surprisingly well. Thus,
our determination of the background light seems to be a reasonable estimate
of the background luminosity density across the entire region studied.
We adopt the background values from 
the number density profile, which are intermediate between
the values for the central region and the Century Survey. 

This estimate of the background is likely
an overestimate. It is conservative in the sense that it tends to
underestimate the cluster light. Because the amount of background light
enclosed increases with radius, the effects of overestimating the
background also increase with radius.

\section{Surface Number Density Profile}

One approach to studying the relative distributions of mass and
light in a cluster is to compare the infall mass profile with the
galaxy surface
number density profile. We calculate the surface number density profiles
with two different galaxy samples $m_{16.5}$ and $m_{18}$
where the subscript denotes the magnitude limit of the sample. For both
samples, we exclude non-members with 
$m_R<$16.5. The $m_{16.5}$ sample provides a lower limit on the asymptotic
number density for fainter limits. We subtract a constant surface
number density from the $m_{18}$ sample using the method described in
$\S 5$.

If light traces mass, the surface number density profile should
closely resemble the infall mass profile ($\S 3$). The
projection of the NFW profile yields a surface density profile $\Sigma
(R_p)$
\beqn
\Sigma _{NFW} (\tilde{R}) = \frac{N_c}{\pi {\rm ln}(4/e)
a_{NFW}^2 (\tilde{R}^2 -1)} [1-X(\tilde{R})]
\eeqn
where $\tilde{R}=R/a_{NFW}$ is the projected radius in units of the
core radius, $N_c$ is the number of galaxies within $a_{NFW}$, and
\beqn
X(\tilde{R}) = \frac{{\rm sec}^{-1}{\tilde{R}}}{\sqrt{\tilde{R}^2-1}}
\eeqn
Similarly, the projection of the Hernquist profile can
be written as
\beqn
\Sigma _{Hern} (\tilde{R}) = \frac{2 N_c}{\pi
a_{Hern}^2 (\tilde{R}^2 -1)^2} [(2+\tilde{R}^2)X(\tilde{R}) - 3]
\eeqn
(\cite{andi}).

\begin{figure} 
\figurenum{16}
\label{fignprofile}
\plotone{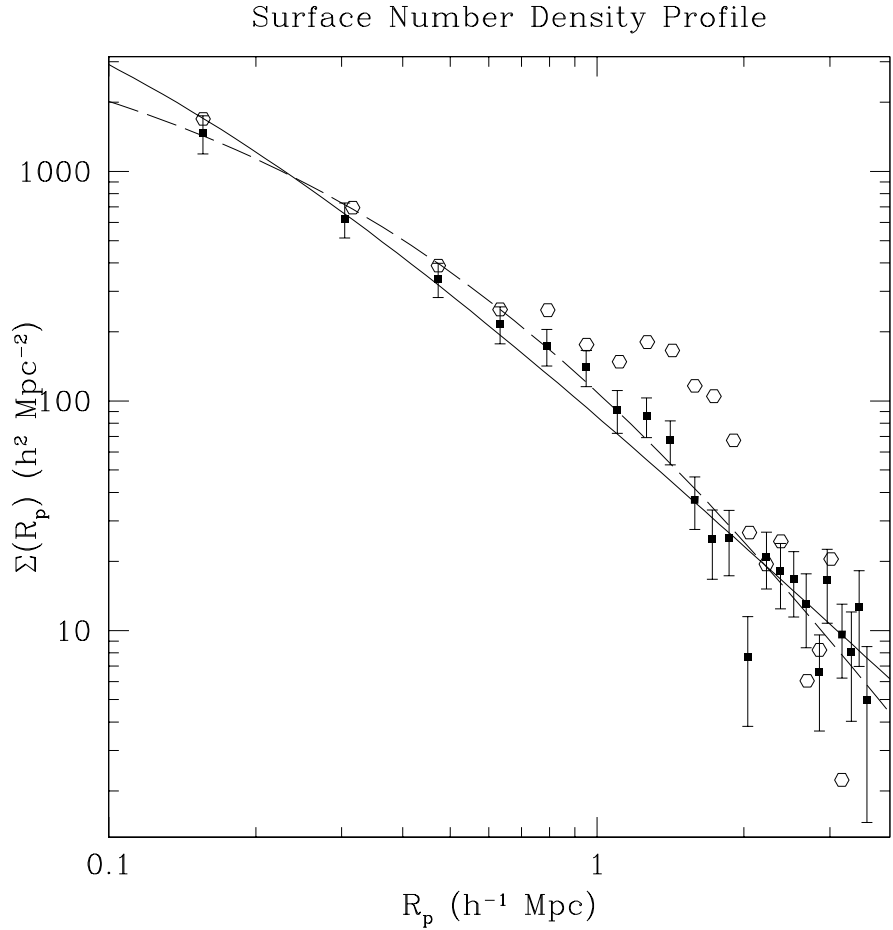} 
\caption{Surface number density profile for A576. Squares show the
profile calculated from 
$m_R<16.5$, hexagons show the background-subtracted profile for
$m_R<18$. The solid and dashed lines are the best fit NFW and
Hernquist profiles respectively for the $m_R<16.5$ sample.}
\end{figure}

Table \ref{nproftable} gives the results of our fits. For the NFW
profile, the scale radius of the surface number density profile for
the $m_{16.5}$ sample agrees with that for the mass profile. The
scale radii for the $m_{18}$ sample and for the Hernquist fit to the
$m_{16.5}$ sample are larger 
than the best-fit scale radii for the respective mass profiles.
We display the surface number density
profiles and the best-fit models in Figure \ref{fignprofile}.
There appears to be a relative excess of galaxies in the magnitude
range $16.5<m_R<18.0$ between 1 and 2$\Mpc$. This excess leads to
significantly larger scale
radii for the $m_{18}$ sample than for the $m_{16.5}$ sample. This
discrepancy may
suggest the presence of 
a background cluster of galaxies ($\S 5$).
The surface number density profile yields some evidence that galaxies
are more extended than the dark matter, just as hot gas in the core is
more extended than the dark matter. This evidence is stronger for the
Hernquist model than for the NFW model; note that the Hernquist models
produce better fits. One possible explanation of this
difference is luminosity segregation, which we investigate next.

\section{Luminosity Segregation}

%We divide the infall region of A576 into two radial bins. We refer to
%these bins, 0-1.45 and 1.45-4$~\Mpc$, as the inner and outer sample
%(we motivate this 
%separation, which corresponds to $\sim 1$ Abell radius, below). 
To convert to absolute magnitudes, we assume that all
galaxies are at the distance of the cluster center; thus,
$M_R = m_R - 35.29 - A_R + 5~{\rm log}~h$, where we estimate the dust
extinction coefficient $A_R$ from the dust maps of Schlegel,
Finkbeiner, \& Davis (1998) as listed in Table 1. We obtain the color
excess at each galaxy position; the inferred extinction values range
from 0.13 to 0.26 across the field. 

\begin{figure} 
\figurenum{17}
\label{figlumfns}
\plotone{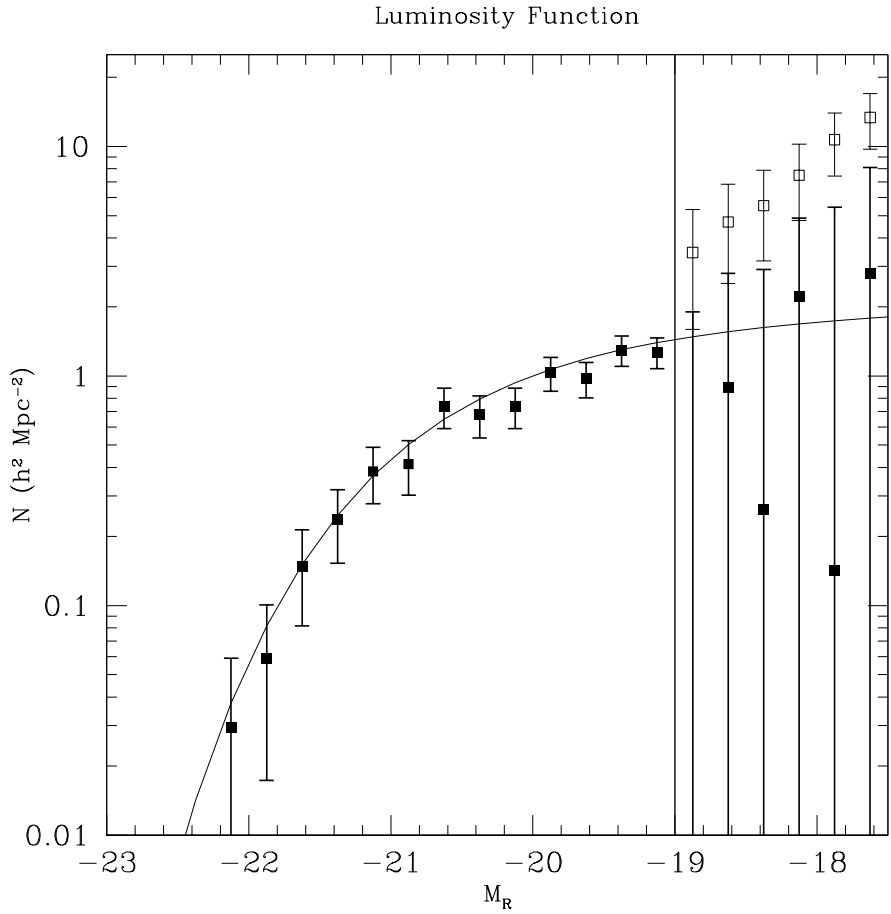} 
\caption{Luminosity function for entire region. Filled and open
squares are calculated with and without background subtraction
respectively. The vertical line is the limit of our survey. The curve
indicates the best-fit Schechter luminosity function for $M_R<-19$.}
\end{figure}

Figure \ref{figlumfns} shows the luminosity distribution for all
galaxies with $m_R<18$ after removing all non-cluster galaxies with 
$m_R<$16.5, the completeness limit of our spectroscopic sample. The
contribution of background galaxies steepens the distributions at 
$M_R \gtrsim -19$. In the outer regions, the number counts begin to
resemble $N(m) \propto m^{0.6}$.  

\begin{figure} 
\figurenum{18}
\label{figlumvsr}
\plotone{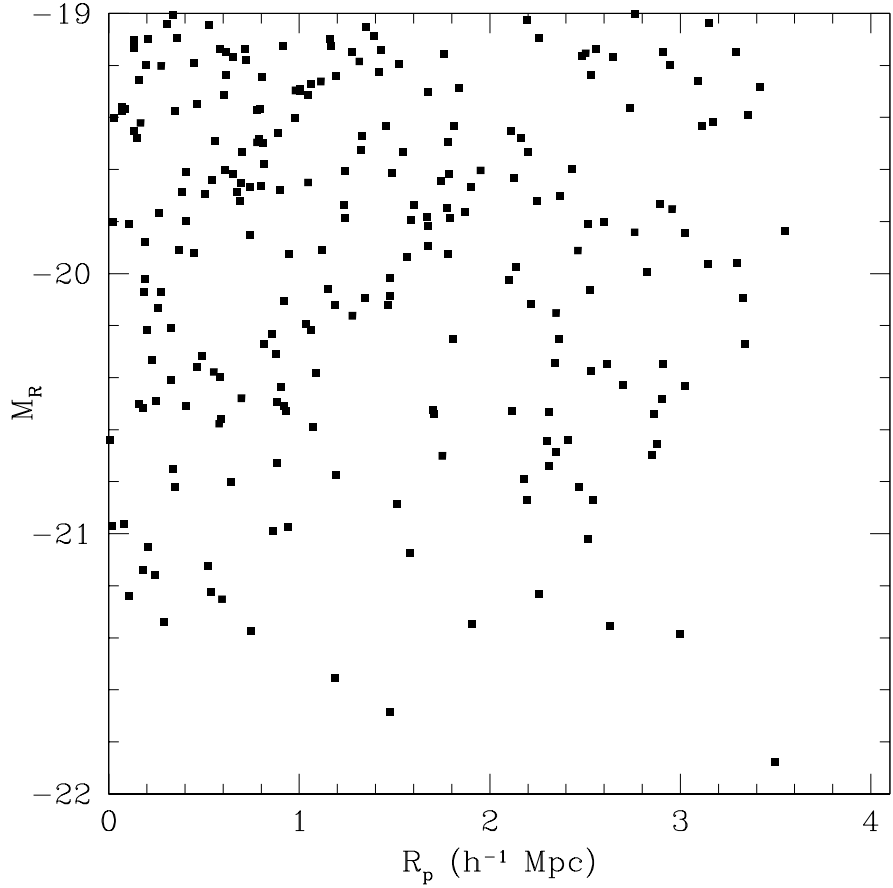} 
\caption{Absolute magnitude versus projected radius. Significant
luminosity segregation requires that absolute magnitudes increase with
radius. The three most luminous galaxies all have $R_p > 1~\Mpc$.}
\end{figure}

Luminosity segregation, if present, should be most apparent
for the most luminous galaxies. In a large 
sample of clusters, Adami et
al.~(1998a) claim luminosity segregation among (on average) the four
brightest members of each cluster.
As a simple test of luminosity segregation, Figure \ref{figlumvsr}
shows absolute magnitude versus radius for cluster galaxies in the
complete $m_{16.5}$ sample. Significant segregation requires increasing 
absolute magnitude with radius. 
In fact, the three brightest galaxies in A576 are all $>
1~\Mpc$ from the center, and the brightest cluster galaxy is
$3.5~\Mpc$ from the center, more than two Abell radii away. 
 The most luminous galaxy appears to be a
normal S0 galaxy, with an absorption dominated spectrum and no close
companions. The next two 
most luminous galaxies also have absorption dominated spectra, but
one of them has a close companion. 

\begin{figure} 
\figurenum{19}
\label{figlumfnsks}
\plottwo{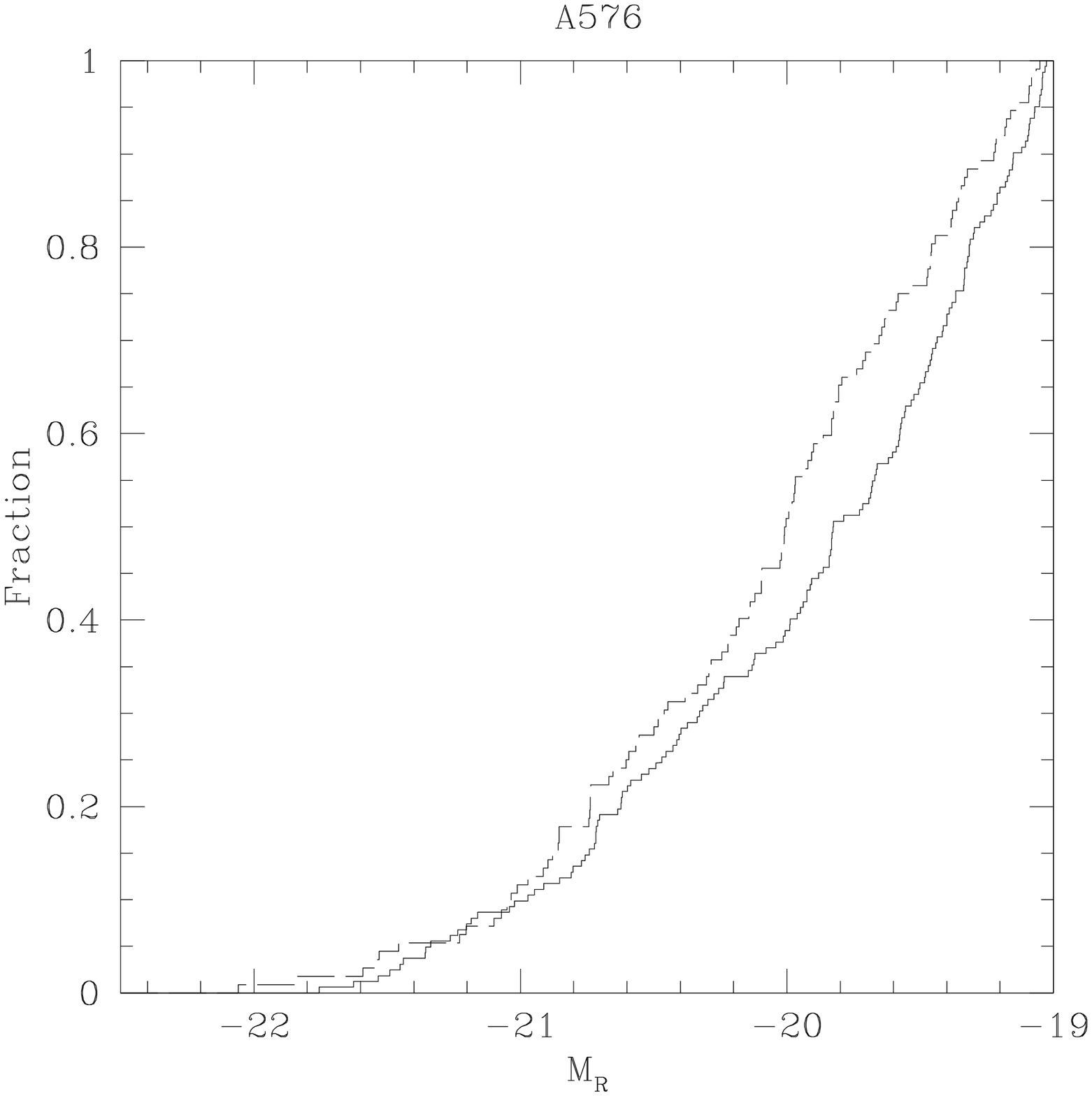}{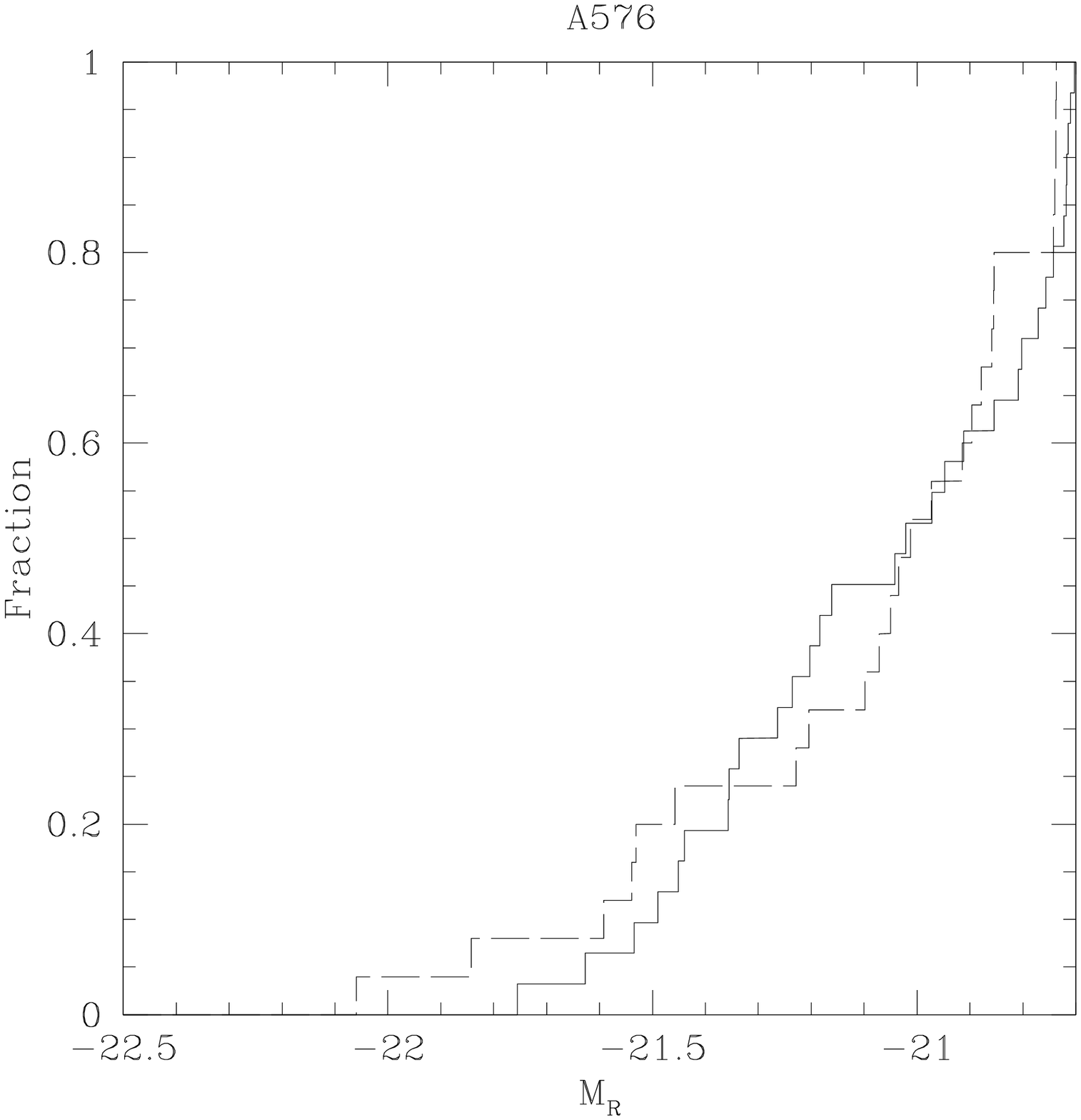} 
\caption{(a) Cumulative absolute magnitude distributions of galaxies
inside (solid lines) and outside (dashed lines) $R_{div} = 1.45~\Mpc$. The
outer sample is brighter than the inner with 95\% confidence for
$M_R<-19.0$. (b) Same for $R_{div} = 1.45~\Mpc$ and $M_R<-20.7$.}
\end{figure}

As a quantitative test of luminosity
segregation, we compare the cumulative luminosity distributions for all 
cluster galaxies with $M_R<-19.0$ (our spectroscopic completeness
limit) inside and outside a radial cutoff 
(e.g., Figure \ref{figlumfnsks}). We vary this cutoff radius from
$0.05-2.50~\Mpc$ in steps of $0.05~\Mpc$ and perform a K-S test for each
division. The only radial division that yields a separation at greater
than a 90\% confidence level is at $1.45~\Mpc$ (95.2\%
confidence). However, the difference between the samples suggests
luminosity anti-segregation, i.e., the outer sample is marginally
brighter than the inner sample.
 
To test the sensitivity of this result to the magnitude
limit, we repeat the analysis using a magnitude cutoff of $M_R<-20.7$,
approximately the value of $M_*$ for the entire region ($\S
8.1$). With this cutoff, there is no significant difference at any
radius (see Figure \ref{figlumfnsks}).  
Combined with the result that the three most luminous galaxies in
the sample are located far outside the core, we conclude that there is
no significant evidence of luminosity segregation in A576.

\section{Luminosity Function and Light Profile}

\subsection{Luminosity Function}

By correcting the luminosity function for the background counts ($\S 5$),
we obtain the luminosity function shown in Figure \ref{figlumfns}. 
Because we explicitly assume that few faint member galaxies are present at
large radii, we are unable to constrain the faint-end slope of the
luminosity function.

We calculate the best-fit Schechter (1976) luminosity function,
\beqn
\frac{dn}{dM} \Bigg{\vert}_M \propto 10^{0.4(1+\alpha)(M_*-M)} {\rm exp} [-
10^{0.4(M_*-M)}]
\eeqn
in the range $-22.5<M_R<-19$, where $M_*$ is the characteristic
absolute magnitude and $\alpha$ is 
the slope at the faint end.  We find the best-fit form by
minimizing $\chi^2$. The parameters for the total luminosity function,
$M_* = -20.7\pm0.4$ and $\alpha = -1.0\pm0.3$ (95\% confidence limits),
are consistent with those of M96, LCRS (\cite{lcrslfn}, a hybrid
Gunn $r$-Kron-Cousins $R$ system), and the Century Survey (\cite{cslfn}, an
R-selected survey). The fit yields $\chi ^2 = 3.5$ for 10 degrees of
freedom; a single Schecter function is therefore a good model of the
data. 

\subsection{Light Profile}

\begin{figure} 
\figurenum{20}
\label{figlightprofile}
\plotone{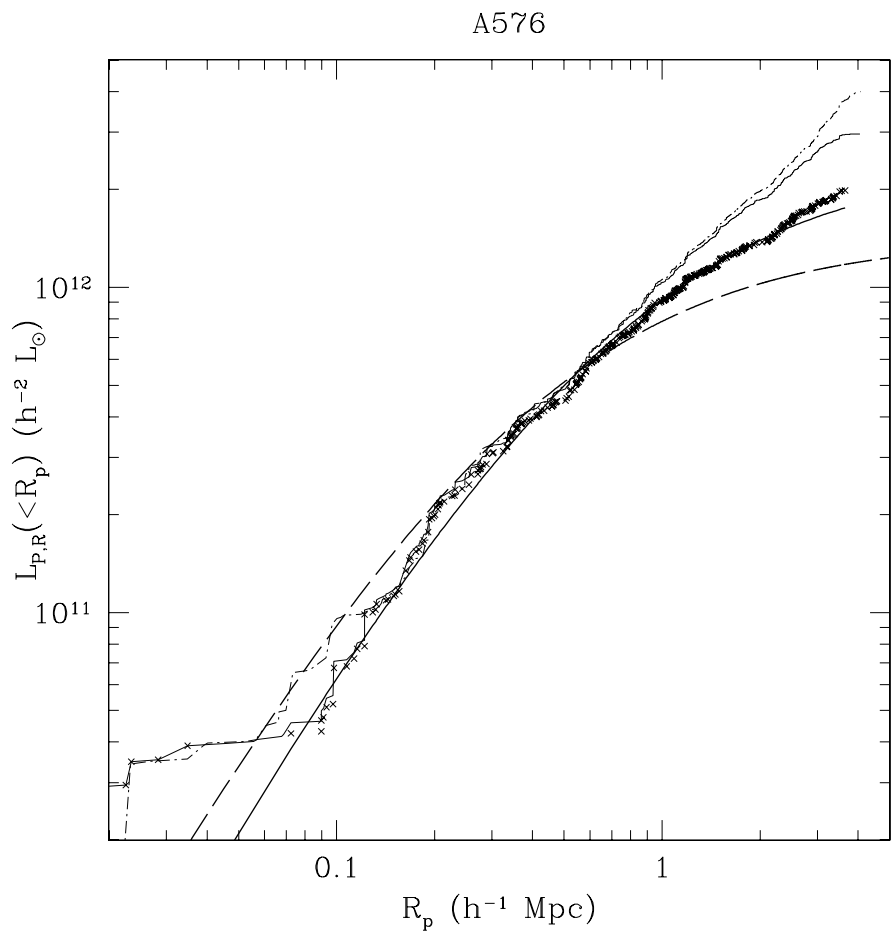} 
\caption{R-band luminosity profile of A576 (crosses). We exclude all
non-cluster galaxies with $m_R<16.5$. The faint solid
and dash-dot lines are the profiles
calculated from $m_R<18$ galaxies with and without background
subtraction respectively. The heavy solid and dashed lines are the
best-fit Hernquist light and mass profiles (converted with an arbitrary
mass-to-light ratio).}
\end{figure}

We calculate the light profiles for the $m_{16.5}$ and
$m_{18}$ using the estimated background light from the asymptotic
number counts around A576 (Figure \ref{figlightprofile}). Using the
absolute R-band magnitude of the Sun ($M_R = 4.3$ ; Zombeck 1990), we
calculate the total light profiles for these subsamples.
The potential overestimate of the background leads to an underestimate of 
the light profile, particularly at large radii. The $m_{18}$ sample
yields an upper limit to the light profile  by assuming that all
galaxies without redshifts are cluster members (a lower limit comes
from the $m_R<16.5$ sample).

\begin{figure} 
\figurenum{21}
\label{figlproffit}
\plotone{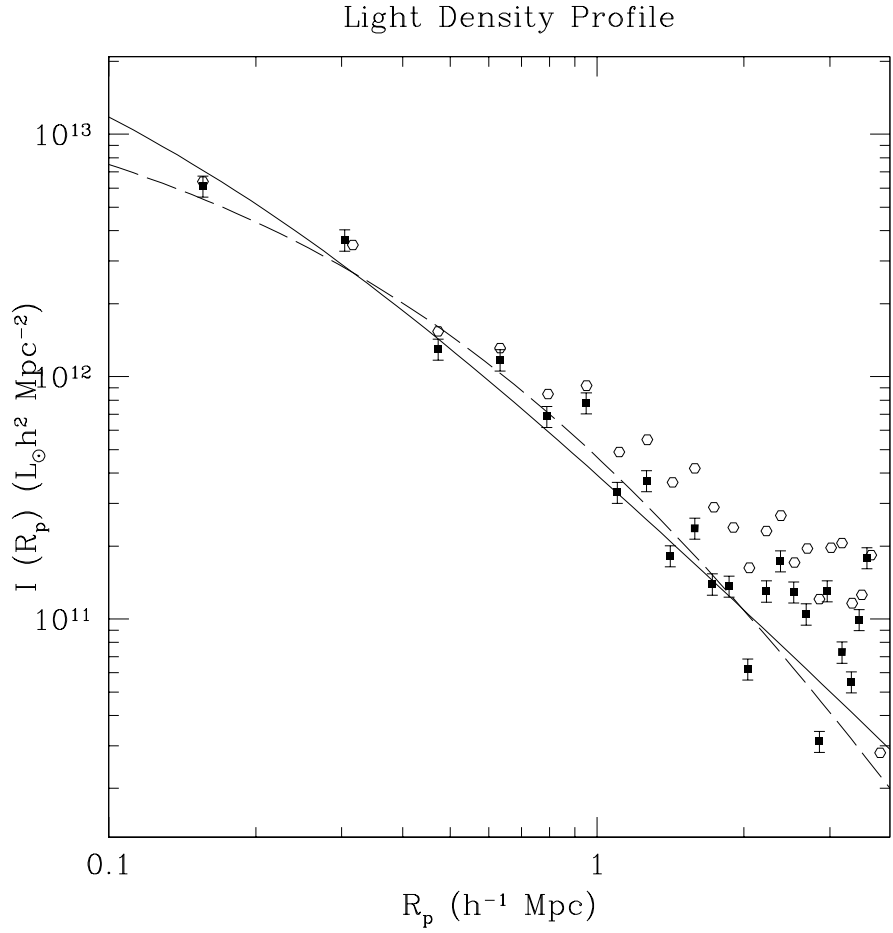} 
\caption{R-band luminosity density profile of A576. Symbols are as in
Figure \ref{fignprofile}. The solid and dashed lines are the best fit NFW and
Hernquist profiles respectively for the $m_R<16.5$ sample.}
\end{figure}

Because the line-of-sight distribution of the galaxies is unknown, the
observed light profile is the projection of the actual light
profile. We therefore compare the observed light profile to the
analytic forms 
in $\S 5$, replacing $N_c$ with $L_c$, the amount of light enclosed
within the scale radius $a$. The cumulative projected light
profile for the Hernquist model has the simple form 
\beqn
\label{projhern}
L(<R_p) = L_{tot} a^2 \frac{X(\tilde{R})-1}{1-\tilde{R}^2}
\eeqn
where $L_{tot}$ is the total luminosity of the system (\cite{h90}). The
results, shown in Table \ref{lprof}, reveal that the scale radii of
the best-fit NFW profiles agree with that of the infall mass profile,
whereas the Hernquist light profiles have  scale
radii a factor of $\sim 1.8-3$ larger than the best-fit mass
profile. Figure \ref{figlproffit} 
displays the best-fit profiles of each type. The values of $\chi ^2$
are rather large in all cases. It is evident from inspection of Figure
\ref{figlproffit} that the observed light profile does not fall off as
steeply as the NFW or Hernquist profiles at large radii ($R_p \gtrsim
2\Mpc$). If we convert the projected mass profile into a projected
light profile by assuming a constant mass-to-light ratio (Figure
\ref{figlightprofile}), the predicted light profile
increases more slowly than the observed light profile. This result
suggests that the difference in scales between the mass and light
profiles can not be explained by projection effects unless there are
significant departures from spherical symmetry (e.g., our line of
sight is perpendicular to the disk of an oblate system).

\section{Mass-to-Light Profile}

\begin{figure} 
\figurenum{22}
\label{figmasstolight}
\plotone{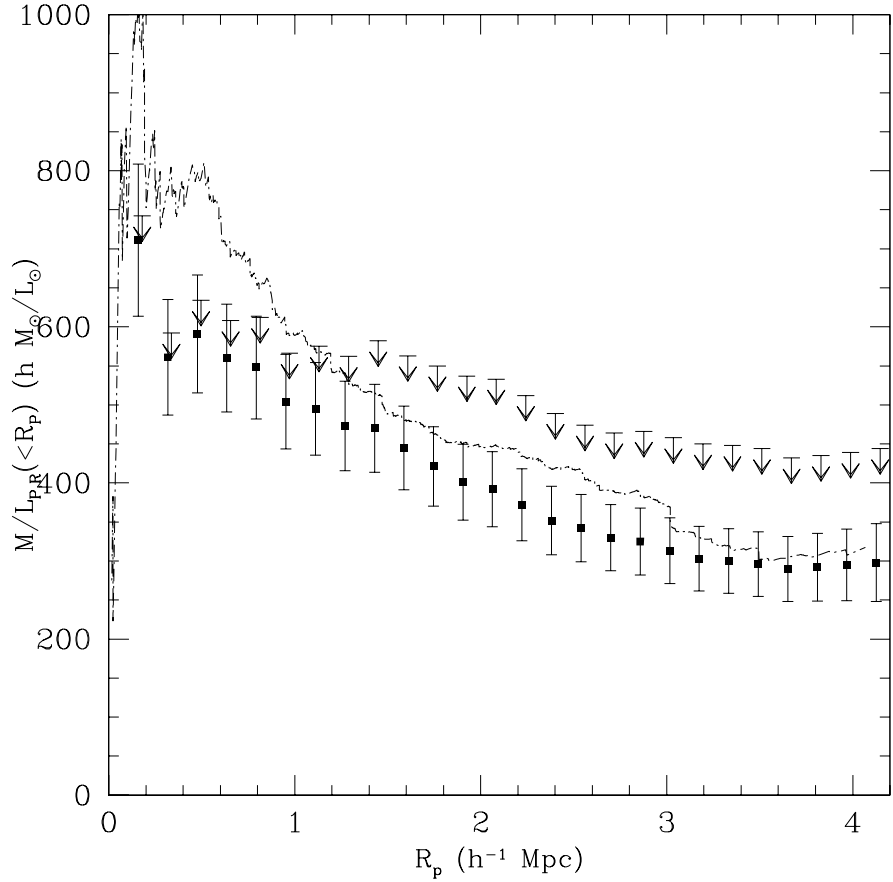} 
\caption{Cumulative mass-to-light profile of A576 calculated from
the infall mass profile and the projected light profile. Squares
estimate the light profile from galaxies with $m_R<18$; upper
limits include only light from confirmed members
($m_R<16.5$). The dash-dotted line indicates the projected best-fit
Hernquist mass profile divided by the projected light profile.}
\end{figure}

With the independently derived mass and light profiles, we can now
calculate the R-band mass-to-light profile of A576 out to $\sim
4~\Mpc$, the limit of the infall mass profile.  
By dividing the mass profile by the projected light
profile, we obtain the mass-to-light profile shown in Figure
\ref{figmasstolight}. 
$M/L_R (<R_p)$ is largest in the core 
($\sim 700~h$) and decreases smoothly to the limiting radius of 
$\sim~4~h^{-1}$~Mpc, where $M/L_R \sim 300 h$. We use the
lower limits on the light profile to calculate ``upper limits'' on
$M/L_R (<R_p)$. These upper limits do not include the uncertainties in
the mass profile; they demonstrate that the decreasing
mass-to-light profile is not a result of underestimating 
the background light (in $\S 5$, we explain why we may overestimate
the background light). We fit the profiles from the 
$(m_{18})$ sample to a straight line ($M/L_R = a R_p + b$) and to a
constant value ($M/L_R = b$) using
weighted least squares.  Table \ref{mlfits} displays the results of these
fits. The best-fit straight lines have negative slopes with high
significance. Because the true $M/L$ is always positive, the
actual $M/L$ profile extrapolated to arbitrarily large radius cannot
be a straight line with a negative slope. Again, because the values of
$M(<R)$ are not independent, the values of $\chi ^2$ are only
indicative. However, this exercise shows that a decreasing profile
yields a much better fit than a flat one.

We note again that the light profile is
the two-dimensional projection of the three-dimensional light profile; 
the deprojected mass-to-light profile of a spherically symmetric
system would decrease more steeply than the ones presented here (see
also Figure \ref{figlightprofile}). As an example, we take the
best-fit Hernquist mass profile, project it according to Equation
\ref{projhern} (assuming spherical symmetry), and divide it by the
projected light profile. The resulting projected-mass to
projected-light profile, shown in Figure \ref{figmasstolight}, is
larger in the inner $\sim 1\Mpc$ than the radial-mass to
projected-light profile. 

\begin{figure} 
\figurenum{23}
\label{figdiffmasstolight}
\plotone{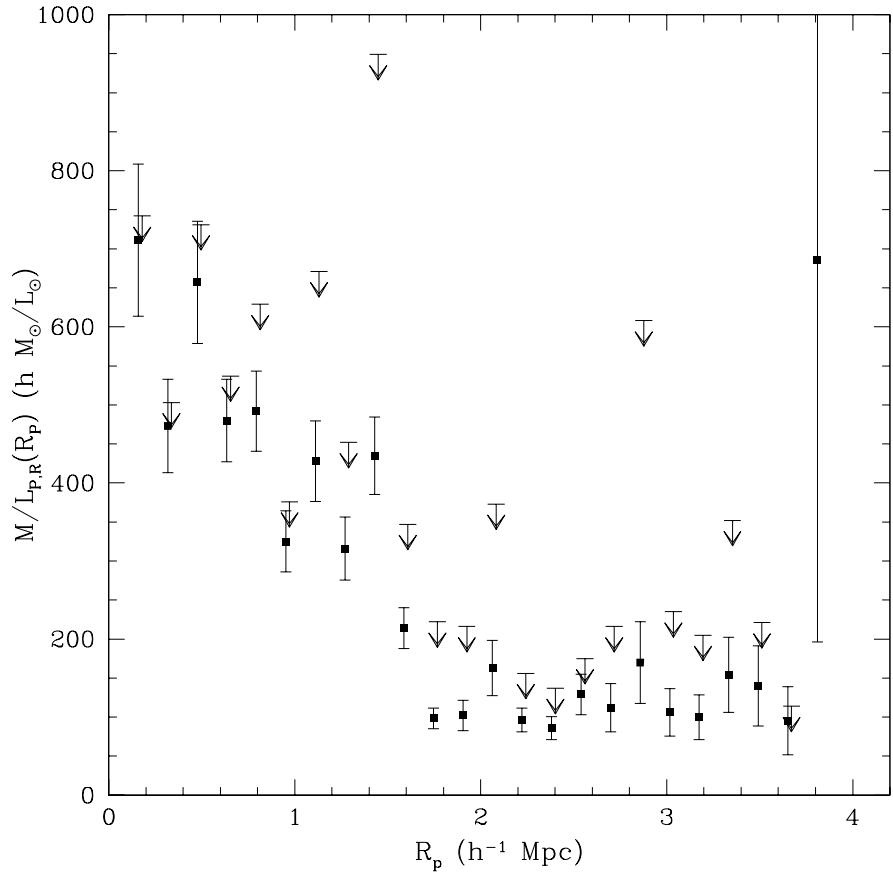} 
\caption{Differential mass-to-light profile. Squares
represent the $m_{18}$ luminosity profile. Upper limits 
reflect the minimum light associated with the cluster.}
\end{figure}

This mass-to-light profile is an integrated profile, i.e., $M/L_R
(<R_p)=~M(<R_p)/L(<R_p)$. The significance of the decreasing profile
is more dramatic as a differential profile where
$M/L_R (R_p) = \delta M(R_p)/\delta L(R_p)$ (Figure \ref{figdiffmasstolight}). 
Again, a decreasing mass-to-light profile yields a better fit than a
flat one (Table \ref{mlfits}). The outermost value of this profile ($\sim
100~h$) should be  
closest to an estimate of the universal value of $M/L_R$. Interestingly,
this value agrees with the mass-to-light ratios of some elliptical galaxies
(\cite{m94}; \cite{bld95}). 

\section{Discussion}

A decreasing mass-to-light profile is perhaps surprising. One immediately
wonders what systematic effects 
could mimic a decreasing $M/L_R$ profile. Is the mass
profile incorrect? The infall mass profile yields larger values than X-ray
estimates in the inner regions and lower values than virial theorem
estimates in the outer region. Non-thermal pressure support,
nonisothermality, and deviations from virial equilibrium are plausible
explanations for these effects. Another unknown systematic effect is
the actual distribution of galaxy orbits. However, simulations suggest
that this effect is not large (D99). The infall mass profile itself is
uncertain by a factor of 2. Hidden systematic effects could affect the
shape of the mass profile, a possibility we plan to explore with other
systems. The infall mass profile is unstable from
$1.5-2.2 h^{-1}$ Mpc, but beyond this range,  $M/L_R$ continues to
decrease even though the mass profile is well determined. A Jeans
analysis of the virial region will provide an interesting check on the
infall mass profile (\cite{mw00}); preliminary results suggest
good agreement between the infall mass profile and the Jeans mass
profile. We note also that the X-ray mass 
found by White et al.~(1997) provides a lower mass limit at small
radius. While this mass estimate is smaller than the infall mass
estimate, it is not sufficient to reconcile the observations with a
constant mass-to-light ratio.

Is the light profile poorly estimated? In the inner regions, we use
SExtractor magnitudes, which may underestimate the diffuse light
associated with Brightest Cluster Galaxies (BCGs) by $\sim 50\%$
(\cite{gzzd}). Even though 
A576 has no cD galaxy, diffuse light near the center of the cluster
may partially account for the observed decreasing M/L profile. The
magnitude of this effect ($\sim 20\%$ of the total light within
$1.5~\Mpc$), however, is not sufficient to reconcile the 
observed $M/L_R$ profile with a constant value. Background
subtraction always contains some uncertainty; we show the limits of
this uncertainty based on galaxies reliably contained within the
infall region. The background light we assume exceeds that in
randomly selected fields, and two independent estimates of the
background agree remarkably well. The shape of the $M/L_R$ profile is 
insensitive to the magnitude limit. The surface number density profile
also supports a larger scale for the galaxies than for the mass. We
note that any systematic underestimate of total galaxy light (e.g.,
missing halo light) would likely affect the magnitudes of all
galaxies. Such an effect would change the absolute value of $M/L_R$
but would be unlikely to alter the shape of the profile.  

Are projection effects skewing our results? According to the infall
model, the mass profile is a true radial profile; the light
profile is the two-dimensional projection of the actual
light profile. Assuming that luminosity density decreases
monotonically with radius, the projected light profile is steeper than
the actual light profile. This effect would increase the significance
of the decreasing  mass-to-light profiles presented
here. Departures from spherical symmetry could also affect the
mass-to-light profile; looking down the barrel of a prolate spheroid
would exaggerate the decrease in M/L; looking at the disc of an
oblate spheroid would mitigate it. There are no obvious
indications of oblate or prolate structure in the sky distribution of
infall members (Figure \ref{figskyplotz}). Analysis of other systems
should provide an important check on these projection effects.

Is the decreasing mass-to-light profile an effect of calculating the
luminosity from R-band photometry? Galaxies in the cores of clusters
are redder than those far outside the cores. This effect
causes the R-band light density profile to decrease more steeply with
radius than a B-band
light density profile. This effect would oppose the apparent 
trend of a radially decreasing mass-to-light profile. 

M96 demonstrate that the emission-dominated galaxies have different
kinematic properties than absorption-dominated galaxies. On this
basis, M96 removes the light contribution from emission-dominated
galaxies in calculating the mass-to-light ratio. Because many of these
emission-dominated galaxies are in the infall region but probably not
inside the virial radius, the mass-to-light profile is biased low in
the central regions and high in the outer regions.

Finally, we note a curious feature of the D99 simulations. For
$\Lambda CDM$ models, the mass-to-light profiles (in B band relative to
the global value) show a decreasing trend similar to that shown
here. Clusters in a $\tau CDM$ model produce the opposite effect,
mass-to-light profiles which increase with radius. D99 attributes this
effect to the deficit of blue galaxies in the centers of clusters in the
$\Lambda CDM$ model, but later analysis reveals a
similar effect for mass-to-light profiles measured in I band. This
result suggests 
that the decreasing mass-to-light profile may contain information on
the underlying cosmology.

\section{Conclusions}

We calculate the mass profile of Abell 576 to $\sim 4~\Mpc$ from 
the observed infall pattern in redshift space. The amplitude of the
resulting mass profile is larger than determined from 
X-ray observations to their limiting radius of 
0.5 $h^{-1}$ Mpc and smaller than virial mass estimates at larger
radius. The infall mass profile agrees extremely 
well with an NFW profile or a Hernquist (1990) profile, and it is
strongly inconsistent with an isothermal sphere profile. 
 This result agrees well with a similar
analysis of Coma by GDK. Our best-fit mass profile implies
that the fraction of gravitational mass contained in hot gas increases
with radius to the limit of the X-ray data, in agreement with previous
studies of other clusters (e.g., \cite{djf95}; \cite{mv97};
\cite{ef99}) and simulations (\cite{white93}; \cite{e97}). The hot 
gas therefore appears to be more extended than the dark matter.

The decreasing amplitude of the caustics suggests
that the mass increases more slowly than for an isothermal sphere. This
inference is supported by fitting mass profiles to the data. GDK found
that the NFW profile describes the mass profile of 
Coma much more accurately than an isothermal sphere profile, and
Figure 1 of CYE clearly shows the existence of radially decreasing
caustics in the composite CNOC cluster. Many clusters therefore show
evidence of mass profiles shallower than an isothermal sphere; the
masses of clusters probably increase no faster than ${\rm ln}(R_p)$ at
large radius.  

Using photometric data from a
large area surrounding the cluster, we find little
evidence for luminosity segregation. In fact, there is some evidence of
luminosity antisegregation; the three brightest cluster
galaxies lie far outside the center of the cluster. 

The R-band mass-to-light ratio is largest in the core ($\sim 700~h$)
and decreases smoothly to the limiting radius 
$\sim~4~h^{-1}$~Mpc, where $M/L_R (<R_p) \sim 300 h$. The decrease is
more dramatic 
in the differential mass-to-light profile. At the limit of our
sample, $dM/dL \sim 50-150 h$. The luminosity density from the Century
Survey is $\rho_L = \phi^* L^* \Gamma (2+\alpha) = 6.0\pm 2.6 \times 10^8
h^3~$Mpc$^{-3}$ which yields $(M/L)_{crit} = 459 \pm 199 h$
(\cite{cslfn}). Our 
estimate of the global value of $M/L$ thus implies $\Omega _m \lesssim
0.22\pm0.1$, or $\Omega _m \lesssim 0.4$ at a 95\% confidence
level. This estimate of $\Omega _m$ is an upper limit due to the 
unknown contribution of faint ($m_R>18$) galaxies. Note that using the
luminosity function parameters from the LCRS (\cite{lcrslfn}) yields a
significantly smaller luminosity density corresponding to
$(M/L)_{crit} = 1000\pm 150 h$. Adopting this luminosity density reduces our
estimate of the matter density to $\Omega _m \lesssim
0.10\pm0.05$. 
The decreasing $M/L$ profile suggests that the
dark matter is more concentrated than the optical light of the
galaxies. We consider possible systematic effects ($\S 10$) and find
that they generally oppose the decreasing mass-to-light ratio.
If general, our results place strong constraints on
possible variations of mass-to-light ratios with scale as well as on the
global value of $\Omega _m$. 

Photometric observations in other bandpasses would allow an
examination of the effects of stellar populations on the mass-to-light
profile. Similar studies of other clusters, particularly studies of
clusters with better defined infall regions at large radius, 
should test the generality of our results. 

\acknowledgements

This project would
not have been possible 
without the assistance of Perry Berlind and Michael Calkins, the remote
observers at FLWO, and Susan Tokarz, who processed the spectroscopic
data. We also thank Warren Brown, Daniel Koranyi, and Andisheh
Mahdavi for helpful discussions. KR and MJG are supported in part by
the Smithsonian Institution. JJM is supported by Chandra Fellowship
grant PF8-1003, awarded through the 
Chandra Science Center.  The Chandra Science Center is operated by the
Smithsonian Astrophysical Observatory for NASA under contract
NAS8-39073. AD was supported by a Max-Planck-Institut f\"ur
Astrophysik guest post-doctoral fellowship when this work began. We
thank the referee for helpful comments which improved the
presentation of this paper.

\clearpage

%\clearpage

%\clearpage

\clearpage

%\clearpage

\begin{deluxetable}{ccccccc}  
\tablecolumns{8}  
\tablewidth{0pc}  
\label{fastzr}
\tablecaption{Photometric and Spectroscopic Data\tablenotemark{a} \label{fast}}
\small
\tablehead{  
\colhead{}    RA & DEC & $cz$ & $\sigma _{cz}$ &$m_R$ & $\sigma _{m_R}$ & $A_R$ \\
\colhead{}    (J2000) & (J2000) & ($\kms$) & ($\kms$) & &  \\
}
\startdata
 7 10 58.32 & +56 29 25.4 & 39154 & 37 & 15.84 & 0.06 & 0.15 \\ 
 7 10 59.40 & +56 30 15.8 & 14569 & 17 & 16.11 & 0.06 & 0.15 \\ 
 7 11 03.94 & +56 39 38.2 & 16042 & 15 & 16.38 & 0.06 & 0.13 \\ 
 7 11 04.13 & +56 39 46.4 & 15918 & 16 & 15.92 & 0.06 & 0.13 \\ 
 7 11 09.26 & +56 02 13.2 & 13780 & 29 & 15.59 & 0.06 & 0.16 \\ 
\enddata
\tablenotetext{a}{The complete version of this table is in the
 electronic edition of the Journal.  The printed edition contains only
 a sample.} 
\end{deluxetable}

\begin{deluxetable}{lcrr}  
\label{fastznor}
\tablecolumns{4}  
\tablewidth{0pc}  
\tablecaption{Spectroscopic Data\tablenotemark{a} \label{fastnophot}}  
\small
\tablehead{  
\colhead{}    RA & DEC & $cz$ & $\sigma _{cz}$ \\
\colhead{}    (J2000) & (J2000) & ($\kms$)  & ($\kms$) \\
}
\startdata
 6 53 01.56 & +56 15 35.3 & 10057 & 32 \\ 
 6 54 33.82 & +55 14 41.6 & 14215 & 39 \\ 
 6 55 15.14 & +55 28 42.6 & 16899 & 29 \\ 
 6 55 48.17 & +56 04 56.3 & 19347 & 43 \\ 
 6 56 00.38 & +55 24 02.2 & 7908 & 26 \\ 
\enddata	     						  
\tablenotetext{a}{The complete version of this table is in the electronic edition of
the Journal.  The printed edition contains only a sample.}
\end{deluxetable}

\begin{table*}[th] \footnotesize
\begin{center}

\caption{\label{mprof} \sc Mass Profile Fit Parameters}
\begin{tabular}{lcccccc}
\tableline
\tableline

Profile  &$a$ & 95\% & $C$ & 95\% & $\chi ^2$ & $\nu$ \\
& $\Mpc$ & $\Mpc$ & $10^{15} M_\odot /{\rm Mpc}$ & $10^{15} M_\odot /{\rm Mpc}$ & & \\
\tableline 
NFW & 0.13 & 0.11-0.16 & 2.9 & 2.7-3.2 & 7.23 & 25 \\
Hernquist & 0.43 & 0.37-0.49 & 2.5 & 2.3-2.7 & 1.64 & 25  \\
Isothermal & -- & -- & 0.322 & 0.308-0.334 & 179 & 26 \\
\tableline
\end{tabular}
\end{center}
\end{table*}

\begin{table*}[th] \footnotesize
\begin{center}
\caption{\label{nproftable} \sc Surface Number Density Profile}
\begin{tabular}{llcccccc}
\tableline
\tableline

Profile & Sample & $N_c$ & $a$ & 95\%& $\chi ^2$ & $\nu$ & Prob \\
\tableline 
NFW & $m_{16.5}$ & 124 & 0.12 &0.08-0.18 & 33.7 & 20 & 0.03 \\
NFW & $m_{18}$ & 340 & 0.42 & 0.24-0.74 & 10.4 & 20 & 0.96 \\
Hernquist & $m_{16.5}$ & 605 & 0.68 &0.58-0.82 & 28.9 & 20 & 0.09 \\
Hernquist & $m_{18}$ & 1060 & 1.20 & 0.80-1.64 & 11.5 & 20 & 0.93 \\
\tableline
\end{tabular}
\end{center}
\end{table*}

\begin{table*}[th] \footnotesize
\begin{center}

\caption{\label{lprof}\sc Projected Light Profile}
\begin{tabular}{lccccccc}
\tableline
\tableline

Profile & Sample &$a$ & 95\% & $L_c$ & 95\% & $\chi ^2$ & $\nu$  \\
& & $\Mpc$ & $\Mpc$ & $10^{10} L_\odot$ & $10^{10} L_\odot$ & &  \\
\tableline 
NFW & $m_{16.5}$ & 0.15 & 0.12-0.18 & 59 & 56-62 & 333 & 22 \\
NFW & $m_{18}$ & 0.45 & 0.34-0.59 & 152 & 124-188 & 17.9 & 22 \\
Hernquist & $m_{16.5}$ & 0.76 & 0.70-0.84 & 260 & 252-272 & 347 & 22  \\
Hernquist & $m_{18}$ & 1.27 & 1.03-1.59 & 470 & 400-560 & 26.8 & 22  \\
\tableline
CumHern & $m_{16.5}$ & 0.86 & 0.85-0.87 & 245 & 243-248 & 2524 & 365 \\
CumHern & $m_{18}$ & 1.47 & 1.45-1.48 & 459 & 456-463 & 6121 & 1846  \\
\tableline 
\end{tabular}
\end{center}
\end{table*}

\begin{table*}[th] \footnotesize
\begin{center}
\caption{\label{mlfits} \sc Fits of M/L Profile to $M/L_R = a R_p + b$}
\begin{tabular}{cccccc}
\tableline
\tableline

Type & Sample & a & b & $\chi ^2$ & $\nu$ \\
\tableline 
Int & $m_{16.5}$ & -62$\pm 4$ & 651$\pm 10$ & 2.0 & 22 \\
Int & $m_{18}$ & -90$\pm 5$ & 592$\pm 13$ & 5.5 & 22 \\
\tableline
Int & $m_{16.5}$ & -- & 516$\pm 14 $ & 25 & 23 \\
Int & $m_{18}$ & -- & 378$\pm 10$ & 81 & 23 \\
\tableline 
Diff & $m_{16.5}$ & -158$\pm 31$ & 583$\pm 67$ & 127 & 22 \\
Diff & $m_{18}$ & -109$\pm 27$ & 376$\pm 61$ & 206 & 22 \\
\tableline
Diff & $m_{16.5}$ & -- & 263$\pm 10 $ & 278 & 23 \\
Diff & $m_{18}$ & -- & 145$\pm 6$ & 354 & 23 \\
\tableline
\end{tabular}
\end{center}
\end{table*}


\begin{thebibliography}{}
%\bibitem[Abraham et al.~1996]{ab96} Abraham, R. et al.~1996, \apj, 471, 694
\bibitem[Abell 1958]{abell} Abell, G.O. 1958, \apjs, 3, 211
\bibitem[Aceves \& Perea 1999]{ap99} Aceves, H. \& Perea, J. 1999, \aap, 345, 439
\bibitem[Adami et al.~1998a]{abm98a} Adami, C., Biviano, A., \& Mazure,
A. 1998, \aap, 331, 439
\bibitem[Adami et al.~1998b]{enacs98} Adami, C., Mazure, A., Katgert, P. \& Biviano, A. 1998, \aap, 336, 63
\bibitem[Anupama et al.~1994]{m67a} Anupama, G.C., Kembahvi, A.K.,
Prabhu, T.P., Singh, K.P., \& Bhat, P.N. 1994, A\&AS, 103, 315
\bibitem[Bahcall, Lubin, \& Dorman 1995]{bld95} Bahcall, N.A., Lubin,
L.M., \& Dorman, V. 1995, \apj, 447, L81
\bibitem[Bertin \& Arnouts 1996]{sex} Bertin, E., \& Arnouts, S. 1996,
A\&AS, 117, 393
%\bibitem[Blumenthal et al.~1984]{bfpr} Blumenthal, G.R., Faber, S.M.,
%Primack, J.R., \& Rees, M.J. 1984, \nat, 311, 517
\bibitem[Brown et al.~2000]{warren} Brown, W.R. et al. 2000, in preparation
%\bibitem[Carlberg et al.~1997]{c97c} Carlberg, R.G., et al. 1997, \apj, 485, L13
\bibitem[Carlberg, Yee, \& Ellingson 1997]{c97b} Carlberg, R.G., Yee,
H.K.C., \& Ellingson, E. 1997, \apj, 478, 462 (CYE)
\bibitem[Carlberg at al.~1997]{c97a} Carlberg, R.G., Yee, H.K.C.,
Ellingson, E., Morris, S.L., Abraham, R., Gravel, P., Pritchet, C.J.,
Smecker-Hane, T., Hartwick, F.D.A., Hesser, J.E., Hutchings, J.B., \&
Oke, J.B. 1997, \apj, 476, L7
\bibitem[Carlberg et al.~1996]{c96} Carlberg, R.G., Yee, H.K.C.,
Ellingson, E., Abraham, R., Gravel, P., Morris, S., \& Pritchet,
C.J. 1996, \apj, 462, 32 
\bibitem[Cavaliere \& Fusco-Femiano 1976]{cff76} Cavaliere, A. \& Fusco-Femiano, R. 1976, A\&A, 49, 137
\bibitem[Chevalier \& Ilovaisky 1991]{m67b} Chevalier, C. \&
Ilovaisky, S.A. 1991, A\&AS, 90, 225
\bibitem[Churazov et al.~1996]{chur96} Churazov, E., Gilfanov, M., Forman,
W., \& Jones, C. 1996, \apj, 471, 673
\bibitem[David et al.~1990]{david90} David, L.P., Arnaud, K.A.,
Forman, W., \& Jones, C. 1990, \apj, 356, 32
\bibitem[David et al.~1993]{david93} David, L.P., Slyz, A., Jones, C.,
Forman, W., Vrtilek, S.D. \& Arnaud, K.A. 1993, \apj, 412, 479
\bibitem[David, Jones, \& Forman 1995]{djf95} David, L.P., Jones, C.,
\& Forman, W. 1995, \apj, 445, 578
\bibitem[deLapparent, Geller, \& Huchra 1986]{dgh86} de Lapparent, V.,
Geller, M.J., \& Huchra, J.P. 1986, \apj, 302, L1
\bibitem[den Hartog \& Katgert 1996]{dk96} den Hartog, R. \&
Katgert, P. 1996, \mnras, 279, 349
\bibitem[Diaferio et al.~1999]{dkw99} Diaferio, A., Kauffmann, G.,
Corlberg, J.M., \& White, S.D.M. 1999, \mnras, 307, 537
\bibitem[Diaferio 1999]{d99} Diaferio, A. 1999, \mnras, 309, 610 (D99)
\bibitem[Diaferio 2000]{d00} Diaferio, A. 2000, in preparation
\bibitem[Diaferio \& Geller 1997]{dg} Diaferio, A. \& Geller, M.J. 1997, \apj, 481, 633 (DG)
\bibitem[Dressler 1978]{d78} Dressler, A. 1978, \apj, 226, 55
\bibitem[Dressler et al.~1987]{ga} Dressler, A., Faber, S.M.,
Burstein, D., Davies, R.L., Lynden-Bell, D., Terlevich, R.J., \&
Wegner, G. 1987, \apj, 313, L37
\bibitem[Ellingson et al.~1999]{ellingson} Ellingson, E., Lin, H.,
Yee, H.K.C., \& Carlberg, R.G. 1999, astro-ph/9909074
\bibitem[Ettori \& Fabian 1999]{ef99} Ettori, S. \& Fabian, A.C. 1999,
\mnras, 305, 834 
\bibitem[Evrard 1997]{e97} Evrard, A.E. 1997, \mnras, 292, 289
\bibitem[Evrard, Metzler, \& Navarro 1996]{emn96} Evrard, A.E., Metzler, C.A., \& Navarro, J.F. 1996, \apj, 469, 494
\bibitem[Faber \& Gallagher 1979]{fg79} Faber, S.M., \& Gallagher,
J.S. 1979, \araa, 17, 135
\bibitem[Fabricant et al.~1998]{f98} Fabricant, D., Cheimets, P., Caldwell, N., \& Geary, J. 1998, \pasp, 110, 79
\bibitem[Fabricant \& Hertz 1990]{decaspec} Fabricant, D. \& Hertz,
E. 1990, SPIE Proc., 1235, 747
\bibitem[Fabricant, Lecar, \& Gorenstein 1980]{flg80} Fabricant, D.,
Lecar, M., \& Gorenstein, P. 1980, \apj, 241, 552
\bibitem[Fadda et al.~1996]{f96} Fadda, D., Girardi, M., Giuricin, G., Mardirossian, F., \& Mezzetti, M. 1996, \apj, 473, 670
%\bibitem[Garilli et al.~1999]{glfn99} Garilli, B., Maccagni, D., \&
%Andreon, S. 1999, \aap, 342, 408
\bibitem[Geller et al.~1997]{cslfn} Geller, M.J., Kurtz, M.J., Wegner,
G., Thorstensen, J.R., Fabricant, D.G., Marzke, R.O., Huchra, J.P.,
Schild, R.E., \& Falco, E.E. 1997, \aj, 114, 2205
\bibitem[Geller, Diaferio, \& Kurtz 1999]{gdk} Geller, M.J., Diaferio,
A. \& Kurtz, M.J. 1999, \apj, 517, L23
\bibitem[Giovanelli \& Haynes 1985]{gh85} Giovanelli, R. \& Haynes,
M.P. 1985, \aj, 90, 2445
\bibitem[Girardi et al.~1998]{g98} Girardi, M., Giuricin, G.,
Mardirossian, F., Mezzetti, M., \& Boschin, W. 1998, \apj, 505, 74
\bibitem[Girardi et al.~2000]{g99} Girardi, M., Borgani, S., Giuricin, G.,
Mardirossian, F., \& Mezzetti, M. 2000, \apj, 530, 62
\bibitem[Gonzalez et al.~2000]{gzzd} Gonzalez, A.H., Zabludoff, A.I.,
Zaritsky, D., \& Dalcanton, J.J. 2000, astro-ph/0001415
\bibitem[Gotthelf 1996]{ascaposn} Gotthelf, E. The {\em ASCA} Source
Position Uncertainties, {\em ASCA} Newsletter, Issue
4, available at http://heasarc.gsfc.nasa.gov/docs/asca/newsletters/Contents4.html
%\bibitem[Grogin 2000]{norm} Grogin, N.A. 2000, in preparation
\bibitem[Heisler, Tremaine, \& Bahcall 1985]{htb85} Heisler, J.,
Tremaine, S., \& Bahcall, J.N. 1985, \apj, 298, 8
\bibitem[Hernquist 1990]{h90} Hernquist, L. 1990, \apj, 356, 359
\bibitem[Jones \& Forman 1984]{jf84} Jones, C. \& Forman, W. 1984,
\apj, 276, 38
\bibitem[Jones \& Forman 1999]{jf99} Jones, C. \& Forman, W. 1999,
\apj, 511, 65
\bibitem[Kaiser 1987]{kais87} Kaiser, N. 1987, \mnras, 227, 1
\bibitem[Kaiser et al.~2000]{k99} Kaiser, N., Wilson, G., Luppino, G.,
Kofman, L., Gioia, I., Metzger, M., \& Dahle, H. 1999, \apj, submitted
(astro-ph/9809268)
\bibitem[Kauffmann et al.~1999a]{kauf98a} Kauffmann, G., Colberg, J.M.,
Diaferio, A., \& White, S.D.M. 1999a, \mnras, 303, 188
\bibitem[Kauffmann et al.~1999b]{kauf98b} Kauffmann, G., Colberg, J.M.,
Diaferio, A., \& White, S.D.M. 1999b, \mnras, 307, 529
\bibitem[Koranyi 1999]{dk99} Koranyi, D.M. 1999, private communication
\bibitem[Kurtz \& Mink 1998]{kurtz} Kurtz, M.J. \& Mink, D.J. 1998, \pasp, 110, 934
\bibitem[Landolt 1992]{landolt} Landolt, A.U. 1992, \aj, 104, 340
\bibitem[Lilje \& Lahav 1991]{lilje} Lilje, P.B. \& Lahav, O. 1991, \apj, 374, 29
\bibitem[Lin et al.~1996]{lcrslfn} Lin, H., Kirshner, R.P.,
Schechtman, S.A., Landy, S.D., Oemler, A., Tucker, D.L., \& Schechter,
P.L. 1996, \apj, 464, 60
%\bibitem[Lumsden et al.~1997]{llfn97} Lumsden, S.L., Collins, C.A.,
%Nichol, R.C., Eke, V.R., \& Guzzo, L. 1997, \mnras, 290, 119
\bibitem[Mahdavi et al.~1999]{andi} Mahdavi, A., Geller, M.J.,
B\"ohringer, H., Kurtz, M.J., \& Ramella, M. 1999, \apj, 518, 69 
\bibitem[Markevitch \& Vikhlinin 1997]{mv97} Markevitch, M. \&
Vikhlinin, A. 1997, \apj, 491, 467
%\bibitem[Merritt \& Saha 1993]{} Merritt, D. \& Saha, P. 1993, \apj, 409, 75
\bibitem[Metzler et al.~1999]{metz99} Metzler, C.A., White, M.,
Norman, M. \& Loken, C. 1999, \apj, 520, L9
\bibitem[Mohr et al.~1996]{mohr} Mohr, J.J., Geller, M.J., Fabricant,
D.G., Wegner, G., Thorstensen, J., \& Richstone, D.O. 1996, \apj, 470, 724
\bibitem[Mohr, Geller, \& Wegner 1996]{mgw96} Mohr, J.J., Geller, M.J., \&
Wegner, G. 1996, \aj, 112, 1816
\bibitem[Mohr, Mathiesen, \& Evrard 1999]{mme} Mohr, J.J., Mathiesen,
B., \& Evrard, A.E. 1999, \apj, 517, 627
\bibitem[Mohr \& Wegner 1997]{mw97} Mohr, J.J. \& Wegner, G. 1997, \aj, 114, 25
\bibitem[Mohr et al.~2000]{mw00} Mohr, J.J., et al.~2000, in preparation
\bibitem[Moore et al.~1998]{moore} Moore, B., Governato, F., Quinn,
T., Stadel, J., \& Lake, G. 1998, \apj, 499, L5
\bibitem[Mushotzky et al.~1994]{m94} Mushotzky, R.F., Loewenstein, M.,
Awaki, H., Makishima, K., Matsushita, K., \& Matsumoto, H. 1994, \apj,
436, L79
\bibitem[NFW]{nfw97} Navarro, J.F., Frenk, C.S., \& White, S.D.M. 1997, \apj, 490, 493
%\bibitem[Parzen 1962]{parzen} Parzen, E. 1962, Ann. Math. Stat., 33, 1065
%\bibitem[Pearce et al.~1999]{pvirgo} Pearce, F.R., Jenkins, A., Frenk, C.S., Colberg, J.M., White, S.D.M., Thomas, P.A., Couchman, H.M.P., Peacock, J.A., Efstathiou, G., \& The Virgo Consortium. 1999, \apj, 
%521, L99
\bibitem[Pisani 1993]{p93} Pisani, A. 1993, \mnras, 265, 706
\bibitem[Pisani 1996]{p96} Pisani, A. 1996, \mnras, 278, 697
\bibitem[Praton \& Schneider 1994]{praton} Praton, E.A. \& Schneider,
S.E. 1994, \apj, 422, 46
%\bibitem[Press et al.~1992]{numrec} Press, W.H., Teukolsky, S.A.,
%Vetterling, W.T., \& Flannery, B.P. 1992, Numerical Recipes in Fortran
%77, 2nd ed., Cambridge Univ.~Press, Cambridge
\bibitem[Raymond \& Smith 1977]{rs77} Raymond, J.C. \& Smith,
B.W. 1977, \apjs, 35, 419
\bibitem[Reg\"os \& Geller 1989]{rg89} Reg\"os, E. \& Geller, M. 1989, \aj, 98, 755
\bibitem[Rines et al.~2000]{rines2kb} Rines, K., et al.~2000, in preparation
\bibitem[Rothenflug et al.~1984]{roth} Rothenflug, R., Vigroux, L.,
Mushotzky, R.F., \& Holt, S.S. 1984, \apj, 279, 53
\bibitem[Schechter 1976]{lfnform} Schechter, P. 1976, \apj, 203, 297
\bibitem[Schlegel, Finkbeiner, \& Davis 1998]{sfd} Schlegel, D.J.,
Finkbeiner, D.P., \& Davis, M. 1998, \apj, 500, 525
\bibitem[Silverman 1986]{s86} Silverman, B.W. 1986, Density Estimation
for Statistics and Data Analysis. Chapman \& Hall, London
\bibitem[Small et al.~1998]{small98} Small, T.A., Ma, C.P., Sargent, W.L.W., \& Hamilton, D. 1998, \apj, 492, 45
\bibitem[Struble \& Rood 1991]{sr91} Struble, M.F. \& Rood, H.J. 1991,
\apjs, 77, 363 
\bibitem[van Haarlem et al.~1993]{vanh} van Haarlem, M.P., Cay\'on,
L., Gutti\'errez de la Cruz, C., Martinez-Gonz\'alez, E., \& Rebolo,
R. 1993, \mnras, 264, 71 
\bibitem[van Haarlem \& van de Weygaert 1993]{vanh2} van Haarlem,
M. \& van de Weygaert, R. 1993, \apj, 418, 544
\bibitem[Vedel \& Hartwick 1998]{vedel} Vedel, H. \& Hartwick, F.D.A. 1998, \apj, 501, 509
\bibitem[White, Jones \& Forman 1997]{wjf97} White, D.A., Jones, C.,
\& Forman, W. 1997, \mnras, 292, 419
\bibitem[White 2000]{wtp00} White, D.A. 2000, \mnras, 312, 663
\bibitem[White et al.~1993]{white93} White, S.D.M., Navarro, J.F.,
Evrard, A.E., \& Frenk, C.S. 1993, \nat, 366, 429
\bibitem[Zombeck 1990]{z90} Zombeck, M.V. 1990, Handbook of Space
Astronomy and Astrophysics.  Cambridge University Press, Cambridge 
\bibitem[Zwicky 1933]{z33} Zwicky, F. 1933, Helvetica Physica Acta, 6, 10
\bibitem[Zwicky 1937]{z37} Zwicky, F. 1937, \apj, 86, 217
\end{thebibliography}
\end{document}